\documentclass[apj,twocolumn,twocolappendix]{aastex63}
\newcommand{\Ri}{R_{\rm{in}}}
\newcommand{\etatens}{\overline{\eta}}
\newcommand{\alphatens}{ {\overline{\alpha}}_{\rm dyn} }
\newcommand{\Cs}{c_{\rm s}}
\renewcommand{\vec}[1]{\mathbf{#1}}
\newcommand{\tens}[1]{\mathsf{#1}}
\newcommand{\pd}[2]{\frac{\partial #1}{\partial #2}}
\newcommand{\DS}{\displaystyle}

\usepackage{ctable} 
\usepackage{epstopdf}
\usepackage{pbox}
\usepackage{gensymb}
\usepackage{amsmath}
\usepackage{amssymb}
\begin{document}
\title{MHD accretion-ejection: jets launched by a non-isotropic accretion disk dynamo. I.

       Validation and application of selected dynamo tensorial components}

\author[0000-0003-1454-6226]{Giancarlo Mattia}
\altaffiliation{Member of the International Max Planck Research School for Astronomy \& Cosmic Physics at the University of Heidelberg}
\affiliation{Max Planck Institute for Astronomy, Heidelberg, Germany}

\author[0000-0002-3528-7625]{Christian Fendt}
\affiliation{Max Planck Institute for Astronomy, Heidelberg, Germany}

\correspondingauthor{Giancarlo Mattia}
\email{mattia@mpia.de, fendt@mpia.de}

%/////////////////////////////////////////////////////////////////////////
\begin{abstract}
Astrophysical jets are launched from strongly magnetized systems that host an accretion disk surrounding a central object. 
The origin of the jet launching magnetic field is one of the open questions for modeling the accretion-ejection process.
Here we address the question how to generate the accretion disk magnetization and field structure required for jet launching.
Applying the PLUTO code, we present the first resistive MHD simulations of jet launching including a non-scalar accretion
disk mean-field $\alpha^2\Omega$-dynamo in the context of large scale disk-jet simulations.
Essentially, we find the $\alpha_\phi$-dynamo component determining the amplification of the poloidal magnetic field, which is strictly related to the disk magnetization (and, as a consequence, to the jet speed, mass and collimation), 
while the $\alpha_R$ and $\alpha_\theta$-dynamo components trigger the formation of multiple,
anti-aligned magnetic loops in the disk, with strong consequences on the stability and dynamics of the disk-jet system.
In particular, such loops trigger the formation of dynamo inefficient zones, which are characterized by a weak magnetic field, 
and therefore a lower value of the magnetic diffusivity.
The jet mass, speed and collimation are strongly affected by the formation of the dynamo inefficient zones.
Moreover, the $\theta$-component of the $\alpha$-dynamo plays a key role when interacting with a non-radial 
component of the seed magnetic field. 
We also present correlations between the strength of the disk toy dynamo coefficients and the dynamical 
parameters of the jet that is launched.
\end{abstract}

%//////////////////////////////////////////////////////////////////////////////////////
%/////////////////////////////////////////////////////////////////////////////////////

\section{Introduction}
Astrophysical jets, consisting of collimated high-speed outflows, are launched from a wide range of astrophysical 
objects such as young stellar objects (YSO), micro-quasars or active galactic nuclei (AGNs).
Although these sources span orders of magnitude in term of extension, time scales and energy scales, 
it is commonly accepted that these jets are launched from systems that host an accretion disk surrounding a 
central object \citep{2014prpl.conf..451F, 2015SSRv..191..441H, 2019FrASS...6...54P}.
Further agreement is on the key role of the large-scale magnetic field for the launching, acceleration and 
collimation of jets.
With launching we denote the transition between accretion and ejection, respectively the mass loading
of outflows and jets.
Quite a number of studies have investigated this launching process (see, e.g., 
\citealt{1982MNRAS.199..883B,1985PASJ...37..515U,2002ApJ...581..988C,2006ApJ...651..272F,
2007A&A...469..811Z,2009MNRAS.400..820T,2014ApJ...793...31S,2014ApJ...796...29S,2018ApJ...855..130F}).

Still, the origin of the jet-launching disk magnetic field is not completely understood.
Analytical models and numerical simulations have so far mostly assumed a predefined large-scale,
open magnetic field structure, whose strength and configuration is an essential parameter when
understanding and determining the jet dynamics \citep{2010A&A...512A..82M,2016ApJ...825...14S}.
For the origin of the accretion disk magnetic field one may consider are an extended central stellar 
magnetic field, the advection of magnetic flux from the ambient interstellar medium,
or a magnetic field that is generated by a dynamo process in the disk.
The latter scenario is particularly interesting in order to generate jets from AGN, which 
host as a central object a supermassive black hole, that, unlike stellar objects, cannot produce
its own magnetic field.

Disk dynamos have been suggested already some decades ago
\citep{1981MNRAS.195..881P,1981MNRAS.195..897P,1995ApJ...446..741B}, and
there is a huge literature on dynamo theories and their applications to astrophysics 
(for reviews see e.g. \citealt{2005PhR...417....1B,2019JPlPh..85d2001R}).
Essentially, astrophysical dynamos are thought to be of turbulent, thus small scale nature, while,
on the other hand, one is interested in its large scale effects on the dynamics of these systems.
Overall, it is prohibitively expensive to model the turbulence, involving the smallest scales, and, at the 
same time, aiming to describe astrophysical systems on large scales.
The disk turbulence that may lead to both, a turbulent dynamo effect but also to a turbulent 
magnetic diffusivity is generally thought to be generated by the magneto-rotational instability, MRI \citep{1991ApJ...376..214B}.

For this reason two paths of modelling dynamos have been pursued.
These are (i) direct simulations, which study the natural amplification of the magnetic field (see, e.g., 
\citealt{2010MNRAS.405...41G,2013ApJ...767...30B,2015ApJ...810...59G,2018MNRAS.474.2212R,2018ApJ...861...24H,2020MNRAS.494.4854D}), 
and (ii) the mean-field approach 
(see, e.g., \citealt{1980opp..bookR....K,1995A&A...298..934R,1999MNRAS.310.1175C,2000A&A...353..813R,
2001A&A...370..635B,2006A&A...446.1027C}), 
by which (iia) (semi-)analytical solution can be derived, or (iib) global numerical simulations can be run covering the 
large scales of astrophysical systems.

In this paper we follow the second approach.
Previous work in this field of large-scale disk-jet simulations and a possible origin of large-scale magnetic 
field in accretion disks have been performed by
\citet{2003A&A...398..825V,2014ApJ...796...29S,2018ApJ...855..130F,2018MNRAS.477..127D}, 
essentially demonstrating that a mean-field dynamo generated magnetic field can efficiently launch
jet or outflows.
Most recently, mean-field dynamos have also be considered in general relativistic MHD
tori \citep{BdZ2013, Bugli2014, 2020MNRAS.491.2346T} and disks \citep{Vourellis2020}.

In particular, we extend the work of \citet{2014ApJ...796...29S} by studying the effects of a
{\em non-scalar} mean-field dynamo.
This has not yet been done for launching simulations.
By prescribing a non-isotropic dynamo, in the induction equation, we are able to the disentangle the 
dynamo effects shown in \citet{2014ApJ...796...29S} and \citet{2018ApJ...855..130F} in terms of the 
different components of the dynamo tensor.

Mean-field MHD theory arises from averaging the small-scale dynamics of a turbulent flow pattern,
that is (in our case) affected by a central gravity and a subsequent rotation pattern, gas pressure gradients,
and Lorentz forces.
Both analytical theory (e.g. \citealt{1993A&A...269..581R}) as well as direct numerical simulations resolving 
the disk turbulence (e.g. \citealt{2010MNRAS.405...41G}) have clearly detected an anisotropic nature of the dynamo tensor.
As demonstrated by previous work the different tensor components have different amplitudes and different 
impact on the magnetic field components.
We thus believe that the anisotropy of the dynamo will have essential impact on the structure of the 
dynamo-generated magnetic field.
This has not been shown before in a numerical simulation of jet-launching disks.

The paper is organized as follows. 
In Section \ref{sec:model} we describe our model setup and the numerical approach. 
In Section \ref{sec:toy} we investigate the different effects of the single components of a vectorial dynamo tensor 
on the jet launching process.
We summarize our paper in Section \ref{Sec:conclusions}.
In the Appendix we define our control volumes and provide test simulations comparing our new code to previous
works.

%==============================================================================
%==============================================================================
%==============================================================================

\section{Model approach}
\label{sec:model}

%------------------------------------------------------------------------------
\subsection{MHD equations}
We solve the time-dependent, resistive MHD equations applying the PLUTO code
\citep{2007ApJS..170..228M} version 4.3, on a spherical grid $(R,\theta,\phi)$ assuming axisymmetry.
We refer to $(r,z,\phi)$ as cylindrical coordinates.
The code integrates and solves numerically the set of MHD conservation laws, 
in particular for the conservation of mass,
\begin{equation}
\pd{\rho}{t} + \nabla\cdot(\rho\vec{v}) = 0,
\end{equation}
where $\rho$ and $\vec{v}$ are, respectively, the plasma density and the flow velocity;
the momentum conservation,
\begin{equation}
    \pd{\rho\vec{v}}{t} + \nabla\cdot\left[\rho\vec{v}\vec{v}
    + \left(P + \frac{\vec{B}\cdot\vec{B}}{2}\right)\tens{I} 
    - \vec{B}\vec{B}\right]
    + \rho\nabla\Phi_{\rm g} = 0,
\end{equation}
where $p$ and $\vec{B}$ denote the gas pressure and the magnetic field, respectively.
The central object of mass $M$ provides the gravitational potential $\Phi_{\rm g} = -GM/R$.
The energy is conserved through the equation
\begin{equation}
    \pd{e}{t} + \nabla\cdot\left[\left(e + P + \frac{\vec{B}\cdot\vec{B}}{2}\right)\vec{v}
              - (\vec{v}\cdot\vec{B}) \vec{B}
              + \etatens\vec{J}\times\vec{B}\right] 
              = \Lambda_{\rm{cool}},
\end{equation}
where the total energy density is defined as
\begin{equation}
    e = \frac{P}{\gamma - 1}
      + \rho\frac{\vec{v}\cdot\vec{v}}{2}
      + \frac{\vec{B}\cdot\vec{B}}{2}
      +  \rho\Phi_{\rm g},
\end{equation}
with the polytropic index $ \gamma = 5/3$.

The electric current density is determined by the Ampere's law $\vec{J} = \nabla\times\vec{B}$.
As shown e.g. by \citet{2000A&A...353.1115C,2000A&A...361.1178C,2007A&A...469..811Z,2013MNRAS.428.3151T} 
cooling may play a role in the jet launching process since both density and velocity are subjected to 
cooling effects. 
For the sake of simplicity, as in \citet{2012ApJ...757...65S,2014ApJ...793...31S}, 
the cooling term is set to be equal to the Ohmic heating, which is, therefore, instantly radiated away.

The magnetic field evolution is determined by the induction equation.
Here, we have implemented into the code a mean-field dynamo term 
\citep{1980opp..bookR....K},
\begin{equation}
    \pd{\vec{B}}{t} = \nabla\times(\vec{v}\times \vec{B}+\alphatens \vec{B}-\etatens\vec{J}),
\end{equation}
following mainly the approach of \citet{2014ApJ...796...29S}.
The tensors $\alphatens$ and $\etatens$ describe the $\alpha$-effect of the mean-field dynamo and the magnetic
diffusivity.

%==============================================================================
%
%
%
%==============================================================================

%------------------------------------------------------------------------------
\subsection{Numerical setup}
As we solve the non-dimensional MHD equations, no intrinsic physical scales are involved.
All the primitive MHD variables, i.e. $\rho$,$\vec{v}$,$P$,$\vec{B}$, as well as the length scale and time 
scale, are normalized to their value at the initial inner disk radius $\Ri$.
Thus, velocities are normalized to $v_{\rm{K,in}}$, corresponding to the Keplerian speed at $\Ri$.
As a consequence, the time unit is given in units of $t_{\rm{in}} = \Ri/v_{\rm{K,in}}$, 
and therefore the quantity
$2\pi t_{\rm{in}}$ corresponds to one revolution at the inner disk radius.
In the following, all times are measured in units of $t_{\rm{in}}$, 
implying that $t = 2000$ (in short) corresponds to $t = 2000\,t_{\rm{in}}$.

The computational domain covers a radial range of $R=[1,100]\Ri$ and an angular range of
$\theta=[10^{-8},\pi/2-10^{-8}]\simeq[0,\pi/2]$.
A stretched grid is applied in the radial direction considering $\Delta R = R\Delta\theta$.
The domain is discretized with a number of $[N_R\times N_\theta] = [512\times128]$ grid cells, 
which allows to resolve the initial disk height $H = 0.2 r$ with 16 cells.

For the resolution study (see Appendix paper II) we have discretized the domain with 
$[N_R\times N_\theta] = [1024\times256]$ and $[N_R\times N_\theta] = [256\times64]$ grid cells, namely 32
and 8 cells per disk height, respectively

Our scale-free simulations may be applied, thus scaled to a variety of jet sources.
We apply the same physical scaling as described previous works
\citep{2007A&A...469..811Z,2009MNRAS.400..820T,2012ApJ...757...65S,2014ApJ...793...31S}.
For an astrophysical scaling of our normalized quantities for typical jet systems
we refer to Table \ref{tbl:normalization}.

\begin{table}
\centering
    \begin{tabular}{ccccc}
    \hline \noalign{\smallskip}
    ~         & YSO       & BD         & AGN        & [units] \\     \noalign{\smallskip}    \hline     \noalign{\smallskip}
    $R_0$     & $0.1 $  & $0.01$    & $20$  & AU     \\
    $M_0$     & 1         & 0.05       & $10^8$     & $M_{\odot}$  \\
    $\rho_0$  & $10^{-10}$& $10^{-13}$ & $10^{-12}$ & $g\;cm ^{-3}$~  \\    \noalign{\smallskip}\hline \noalign{\smallskip}
    $v_0$     & 94        & 66         & $6.7 \times 10^{4}$ & $km\;s^{-1}$  \\
    $B_0$     & 15        & 0.5        & 1000       & $G$       \\
    $t_0$     & 1.7       & 0.25       & 0.5        & ${\rm days}$       \\
  $\dot{M}_0$ & $3\times10^{-5}$   & $2\times10^{-10}$ & $10$ & $M_{\odot}yr^{-1}$       \\
  
  \noalign{\smallskip}\hline\noalign{\smallskip}
    \end{tabular}
    \caption{Typical parameter scales for different sources, in particular Young Stellar Objects (YSOs), Brown Dwarfs (BDs) and Active Galactic Nuclei (AGN)}
\label{tbl:normalization}
\end{table}

For spatial integration we use the piecewise parabolic interpolation method (PPM, see \citealt{2014JCoPh.270..784M}).
Time integration is achieved through a third-order Runge-Kutta scheme, while for the flux computation a Harten-Lax-van Leer (HLL) 
Riemann solver is employed \citep{2009book.123..123}.
To preserve the the solenoidal condition of the magnetic field, the method of Upwind Constrained Transport (UCT, \citealt{2004JCoPh.195...17L}) is applied.
In order to achieve stability we choose a Courant-Friedrichs-Lewy time stepping with  $CLF=0.4<1/\sqrt{\rm{N_{dim}}}$.
This may be a challenge for our diffusive MHD simulations, in particular for high resolution
as the diffusive time step goes as $\tau_{\eta}= (\Delta R)^2/\eta$.

%==============================================================================
%
%
%
%==============================================================================

%------------------------------------------------------------------------------
\subsection{Initial conditions}
\label{sec::init}
The simulations start with a very weak initial seed field, thus with a very low disk magnetization,  defined as the 
ratio between the magnetic pressure and the thermal pressure $\mu_{\rm{in}} = B_{\rm in}^2/ P_{\rm in}= 10^{-5}$ 
measured at the disk mid-plane.
Therefore, the initial structure of the accretion disk can be obtained as a solution of the hydrostatic equilibrium between thermal pressure gradients, gravity and centrifugal force 
\citep{2007A&A...469..811Z, 2014ApJ...793...31S}, neglecting the Lorentz force 
\citep{2014ApJ...796...29S, 2018ApJ...855..130F},
\begin{equation}
    \label{eq::steadystate}
    \nabla P + 
    \rho\nabla\Phi_{\rm g} -
    \frac{1}{R} \rho v_\phi^2 (\vec{e}_R \sin\theta + \vec{e}_\theta\cos\theta) = 0.
\end{equation}
This equation can be solved assuming that all the quantities $X$ scale as power laws, $X = X_0 R^{\beta_X} F_X(\theta)$,
where $X_0$ is the corresponding quantity evaluated at the innermost radius of the disk (mid-plane).
For the sake of clarity we summarize here the pertinent formulas. 
Self-similarity requires that every characteristic speed should scale as the Keplerian velocity, $\propto R^{-1/2}$.
Combining this assumption with a polytropic gas, $P\propto\rho^\gamma$, the power law coefficients are
$\beta_{u_\phi} = -1/2$, $\beta_{\rm P} = -5/2$, and $\beta_\rho = -3/2$ 
(as in the self-similar solution of e.g. \citealt{1982MNRAS.199..883B}).
A key parameter to describe the initial disk structure is the ratio between the isothermal sound speed and the
Keplerian velocity at the disk mid-plane of the inner radius $\epsilon=\Cs/v_\phi\left|_{\theta=\pi/2}\right.$.
For an initially thin disk, $\epsilon=z/r=0.1$, and solving for the $z$-component of Eq.~\ref{eq::steadystate} with 
$\rho_{\rm{in}} = 1$ at the inner disk radius, we obtain
\begin{equation}
   F_{\rm P}=\left[\frac{2}{5\epsilon^2}\left(1-\frac{1}{\sin\theta}\right)+\frac{1}{\sin\theta}\right]^{5/2},
\end{equation}
and where we have chosen $P_{\rm{in}} = 0.01$. 
Following the polytropic relation assumed before, the disk pressure is defined by $F_\rho=F_{\rm P}^{3/5}$.

Following \citet{2014ApJ...793...31S, 2014ApJ...796...29S}, by solving the radial component of
Eq.~\ref{eq::steadystate} we obtain the disk rotation profile.
Outside the disk we define a hydrostatic corona,
\begin{equation}
    \rho_{\rm c} = 
    \rho_{\rm{c,in}} R^{1/(1-\gamma)},
     \,\,\,\,
     P_{\rm c}=\frac{\gamma-1}{\gamma}\rho_{\rm{c,in}}
    R^{\gamma/(1-\gamma)},
\end{equation}
with $\rho_{\rm{c,in}}=10^{-3}\rho_{\rm in}$.
All simulations are initialized with a purely radial magnetic field vanishing outside the disk, defined 
by the vector potential
\begin{equation}
    \vec{B}=\nabla\times A\vec{e}_\phi=
    \nabla \times \left[ \frac{B_{\rm{p,in}}}{r}  \exp \left( -8 \left(z/H\right)^2 \right) \right]\vec{e}_\phi.
\end{equation}
Here, $B_{\rm{p,in}} = \epsilon\sqrt{2\mu_{\rm{in}}}$ defines the strength of the initial poloidal 
magnetic field and $\mu_{\rm{in}} = 10^{-5}$ is the initial magnetization along the disk mid-plane.
The quantity $H = 2\epsilon r$ represents the geometrical disk height, respectively twice the initial 
pressure scale height.

%==============================================================================
%
%
%
%==============================================================================

\subsection{Boundary conditions}
The boundary conditions are identical to those of \citet{2014ApJ...793...31S}.
We report them in this section for convenience.
Along the rotational axis and the equatorial plane the standard symmetry conditions are applied.
The inner radial boundary is divided into two different areas.
One is the area that is suited for disk accretion located at $\theta > \pi/2-2\epsilon$,
the other is the coronal area at $\theta < \pi/2-2\epsilon$, 
where we choose $2\epsilon\approx\arctan(2\epsilon)$.
The boundary condition along the inner radial boundary is essential for stabilizing the corona 
against collapse to the central object.
While $v_\theta=0$ along the inner disk boundary, the radial velocity follows a power law, 
$v_R = v_{\Ri} R^{-1/2} \leq 0$,
where the inequality is imposed in order to enforce the boundary behaving as a {"}sink{"}.
Along the coronal area, we prescribe a constant inflow velocity into the domain $v_{\rm p} = 0.2$ (in units of the Keplerian speed 
at $R_{\rm in}$) in the radial direction, that could be interpreted astrophysically as a stellar wind.

From previous jet formation simulations (see e.g. \citealt{1997ApJ...482..712O}) we expect 
the terminal jet speed to reach the Keplerian velocity at the
inner disk.
For $v_\phi$ we prescribe a power law across the inner boundary (for both the disk and the coronal boundary)
\begin{equation}
    v_\phi=v_\phi\big|_{\Ri, R_{\rm{out}} } R^{-1/2}.
\end{equation}
The boundary conditions for the poloidal magnetic field  along the inner radial boundary obey the
divergence-free condition.
The method of constrained transport requires to define only the $\theta$-component of the magnetic field 
along the boundary, while the radial component is recovered from the Maxwell equations.
At the outer boundaries both $B_\phi$ and $B_\theta$ follow a power law 
\begin{equation}
    B_{\phi,\theta}=B_{\phi,\theta}\big|_{\rm{out}}R^{-1},
\end{equation}
while $B_R$ is again recovered using the solenoidality condition.
This is compatible with a constant gradient condition.
For the $B_\phi$ this implies in particular the conservation of the electric current across the boundary.

Along the inner boundary we prescribe $B_\phi = 0$ towards the coronal region, while we again 
adopt a power-law $\propto R^{-1}$ for the boundary area towards the inner disk.
Along the inner radial disk boundary, we prescribe the poloidal magnetic field inclination,
choosing an angle
\begin{equation}
    \varphi = 70\degree\left[1+\exp\left(-\frac{\theta-45\degree}{15\degree}\right)\right]^{-1},
\end{equation}
where $\varphi$ is the angle of the magnetic field with respect to the disk surface.
Note that here again we solve for the divergence-free condition of the magnetic field, recovering the solution
with the inclination prescribed.

These boundary conditions are slightly different from \citet{2014ApJ...796...29S} and \citet{2018ApJ...855..130F}, 
as we not suppress the advection of magnetic flux from the inner disk towards the axis.
This inner boundary condition is known to be quite critical for the numerical stability of the simulation,
as it is time-dependent, thus implementing a feedback loop from the cells of active domain.
While the boundary condition as defined in the papers mentioned above was chosen because it was found to be 
less prone to numerical instabilities,
for the present paper, we decided to release that boundary condition and allow to advect magnetic flux 
across the boundary towards the axial region.
The advection of flux towards the axis has some impact for the structure of this innermost area,
but does not change the structure and the evolution of the surrounding disk jet which is our major focus.
We also think that the advection of magnetic flux towards the axis is a more physical boundary condition.

Across the inner and outer boundaries both the density and the pressure are extrapolated by a power law,
\begin{equation}
    \rho=\rho\big|_{\Ri, R_{\rm{out}} } R^{-3/2} \qquad  P=P\big|_{\Ri,R_{\rm{out}} } R^{-5/2},
\end{equation}
where $\Ri$ and $R_{\rm out}$ is the inner and the outer radius of the domain.
Along the outer boundaries we apply the standard PLUTO outflow (zero gradient) conditions.
In addition, we still prescribe $v_R$ to be non-positive in the disk region and non-negative in the coronal region.

%----------------------------------------------------------------------------------------
\subsection{The dynamo model}
\label{sec::eqdynamo}
For a thin disk, the non-diagonal components of the mean-field dynamo tensor are negligible.
In our approach we consider the explicit form of the dynamo terms following
\citet{1995A&A...298..934R,2000A&A...353..813R},
\begin{equation}
\label{eq::dynamo}
    \alphatens=(\alpha_R,\alpha_\theta,\alpha_\phi) = -\overline{\alpha}_0 \Cs F_\alpha(z),
\end{equation}
where $\Cs$ is the adiabatic sound speed at the disk mid-plane and $F_\alpha(z)$ is a profile function,
\begin{equation}
    F_\alpha(z) = \left\{\begin{array}{ll}
    \sin\left(\pi\DS\frac{z}{H}\right) & z \leq H \\ \noalign{\medskip}
    0 & z > H
    \end{array}\right.
\end{equation}
\citep{2001A&A...370..635B}, that confines the dynamo action within the accretion disk.
Note that \citet{1995A&A...298..934R} have applied a slightly different profile, namely a linear function 
$F_\alpha(z) = z/H$ in the disk.
We prefer the approach of \citet{2001A&A...370..635B} that effectively avoids the discontinuity at the disk surface
and is thus better suited for a simulation that includes also the disk corona.

As in \cite{2014ApJ...796...29S}, we choose a radial dependence of the dynamo $\alpha\propto R^{-1/2}$, since this
profile follows also the sound speed.
Note, however, that compared to our former simulations, in the present setup the radial profile of the 
dynamo is not necessarily constant in time.
As the sound speed is included in the dynamo tensor, along with the disk sound speed, also the dynamo tensor
is updated every time step.
We will demonstrate that this variation has only a minor impact on the overall evolution of the system.
However, it represents a more consistent approach and is furthermore in agreement with the analytical models
of mean-field dynamo theory \citep{1993A&A...269..581R,1995A&A...298..934R}. 

%==============================================================================
%
%
%
%==============================================================================
\subsection{The diffusivity model}
For the magnetic diffusivity tensor we assume a diagonal structure (as for the dynamo).
We adopt an $\alpha$-prescription as typically applied in our previous work \citep{2014ApJ...796...29S},
\begin{equation}
\label{eq::diffusivity}
    \etatens = (\eta_R,\eta_\theta,\eta_\phi)=\overline{\eta}_0\alpha_{\rm{ss}}\Cs H F_\eta(z),
\end{equation}
where $\Cs$ is the adiabatic sound speed at the disk mid-plane, $H$ is the initial disk pressure scale height
while $\alpha_{\rm{ss}}$ is the dimensionless parameter of turbulence \citep{1973A&A....24..337S}.
Thus, the diffusivity is assumed to be essentially of turbulent nature most probably caused by the MRI
\citep{1991ApJ...376..214B}. 
Again we define a profile function,
\begin{equation}
    F_\eta(z) = 
    \left\{\begin{array}{ll}
      1 & z\leq H \\ \noalign{\medskip}
      \exp\left[-2\left(\DS\frac{z-H}{H}\right)^2\right] & z > H
    \end{array}\right.
\end{equation}
that confines the diffusivity within the disk region.
Note that the magnetic diffusivity, or resistivity, respectively, is motivated here as caused by the 
disk turbulence, thus much stronger than the microscopic value.

In the literature of jet launching simulations (see e.g. \citealt{2019MNRAS.490.3112J})
without a mean-field dynamo, the magnetic diffusivity is 
usually computed as
\begin{equation}
    \etatens = \overline{\eta}_0 v_{\rm A} H F_\eta(z),
\end{equation}
where the two model approaches described above coincide if $\alpha_{\rm{ss}} = \sqrt{2\mu_{\rm D}/\gamma}$
where $\gamma$ is the polytropic index and $\mu_{\rm D}$ is the magnetization computed at the disk midplane.
This model approach, in the following denoted as the {\em standard diffusivity} model, 
is, however, not used in this paper. 
One reason is that we want to avoid the accretion instability to occur \citep{2009MNRAS.392..271C}.
This can be avoided when the feedback between the magnetization and the magnetic diffusivity is chosen stronger than 
$\alpha_{\rm{ss}} \propto\sqrt{\mu}$ (see our previous work \citealt{2014ApJ...793...31S}).
Note that we already have a feedback loop on the magnetic diffusivity, as the growth of the magnetic 
field is naturally related to the mean-field dynamo.

We therefore apply the so-called {\em strong diffusivity} model that we have previously invented 
\citep{2014ApJ...793...31S,2014ApJ...796...29S},
\begin{equation}
\label{eq::ssm}
    \alpha_{\rm{ss}} = \sqrt{\frac{2}{\gamma}}\left(\frac{\mu_D}{\mu_0}\right)^2,
\end{equation}
where $\mu_0 = 0.01$.
Since the initial magnetic field does not intersect the disk mid-plane, for the quantity $\mu_{\rm D}$ we 
calculate the ratio between the {\em average} total magnetic field (vertically averaged at a certain radius) 
in the disk and the gas pressure at the disk mid-plane \citep{2014ApJ...796...29S}.
As demonstrated previously, this approach allows to perform a more stable evolution of the disk-jet 
structure over long simulation time \citep{2014ApJ...796...29S,2018ApJ...855..130F}.

%==============================================================================
%
%
%
%==============================================================================

\subsection{Dynamo number and dynamo quenching}
\label{sec::eq_dynnum}
The dynamo number is {\em the} leading parameter for the evolution of the mean-field dynamo-generated magnetic field.
Sub-critical dynamo numbers support a slow, linear growth of the magnetic field, while super-critical numbers lead to 
a rapid, thus exponential growth of the magnetic field.
This exponential growth must be quenched by a physical model that damps the dynamo action (thus the dynamo number) for 
strong fields by first principles.
We define the dynamo number as
\begin{equation}
\label{eq::dynnum}
    {\cal D} =\DS\frac{\alpha_\phi\Omega H^3}{\eta_{\rm{disk}}^2}
\end{equation}
\citep{1995A&A...298..934R,2000A&A...353..813R}.
Since at $z=H$ the dynamo components vanish, the quantity $\alpha_\phi$ is computed at $z = H/2$.
For the definition of an {\em average} disk diffusivity $\eta_{\rm{disk}}$ we refer to Appendix \ref{sec::integrals} .

This definition of the dynamo number is a product of the azimuthal magnetic Reynolds number
${\cal R}_\Omega=|\Delta\Omega|H^2/\eta_{\rm{disk}}$ based on the shear of the flow $\Delta\Omega$,
and the magnetic Reynolds number ${\cal R}_\alpha = \alpha_\phi H / \eta_{\rm{disk}}$, based on the $\alpha$-effect 
(considering $\alpha_\phi$ as the strongest dynamo contribution in disks).

As demonstrated by \citet{2014ApJ...796...29S}, both the disk orbital velocity and the sound speed at the
disk mid-plane undergo some little variation during the temporal evolution of the system. 
Therefore, for an almost constant diffusivity profile with radius, $\cal D$ would scale almost linearly with the radius. 
We note that this is a rough estimate - as the disk diffusivity does not follow a constant radial profile,
even in quasi-steady state.

The dynamo number also depends on $\alpha_{\rm{ss}}$, 
\begin{equation}
    {\cal D} \propto \alpha_{\rm{ss}}^{-2} \propto \mu_{\rm{disk}}^{-4}
\end{equation}
\citep{2014ApJ...796...29S,2018ApJ...855..130F}.
Note that our diffusivity model leads to a rapid growth of the magnetic diffusivity and, as 
a consequence, to a saturation of the magnetic field strength.
We thus expect the dynamo number to converge to a sub-critical strength - at a given radius - at which the
magnetic field cannot be amplified anymore. 
In order to determine such a critical strength for dynamo action involves to consider the full nonlinear 
evolution of the system, and therefore will be subject of analysis in the following sections.

As previously shown \citep{2014ApJ...796...29S}, our diffusivity model is able to suitably quench the dynamo action, 
preventing an endless amplification of the magnetic field.
This is triggered by the strong feedback of the disk magnetization on the magnetic diffusivity  (through $\alpha_{ss}$).

Although this model prevents the accretion instability\footnote{the accretion instability is the disk mass loss which increases the magnetization which increases the mass loss and so on}  described in \citet{2014ApJ...793...31S}, 
it may lead to un-physically large values of $\alpha_{\rm{ss}}$, if not treated with care.
Other quenching models, as e.g. the standard quenching model discussed in \citet{1993A&A...269..581R,1995A&A...298..934R}, 
lead to the same quenching effect as our diffusive quenching model, while keeping the maximum strength of 
$\alpha_{ss}$ in a physically likely range as discussed in e.g. \citet{2007MNRAS.376.1740K}.

However, in order to be able to evolve the long-term properties of the various dynamo models we choose the
diffusive quenching over the standard quenching. 
The standard quenching was found to be prone to the accretion instability for our setup.
Furthermore, the diffusive quenching works much 
smoother compared to the standard quenching and leads to the same result.
We note that we do not put any lower bounds on the turbulence level $\alpha_{\rm{ss}}$.
This may effect, via the dynamo-alpha $\alpha_0$, the critical dynamo number, above which we expect 
an effective magnetic field amplification.
The study of physically more self-consistent feedback models for dynamo quenching will be subject of our future work.

%==============================================================================
%
%
%
%==============================================================================

\section{A toy model for an anisotropic mean-field dynamo}
\label{sec:toy}
This section aims to disentangle the effects that are physically caused by the different components of the dynamo 
tensor (the three components of a vector in our case) in order to gain a detailed understanding of the 
physical process of field amplification at act.

%==============================================================================
%
%
%
%==============================================================================

\subsection{Anisotropic dynamo and diffusivity coefficients}
Our aim is to generalize the dynamo models applied previously \citep{2003A&A...398..825V,2014ApJ...796...29S, 2018ApJ...855..130F}.
These works applied a scalar (thus isotropic) $\alpha$ coefficient.
Here we apply the anisotropy of the dynamo, assuming that the coefficients $\overline{\alpha}_0$, as described in Section
\ref{sec::eqdynamo}, are not necessarily the same \citep{1995A&A...298..934R}.

%-------------------------------------------------
\begin{table}[t]
\caption{Simulations with the dynamo coefficients of $\overline{\alpha}_0 = (\phi,\psi,\chi)\alpha_0$.
The magnetic diffusivity distribution is the same with $\eta_0=0.165$.
The run time of the simulations is $t_{\rm F}$ in units of $1000$.
}
\centering
\begin{tabular}{llllll}
  \noalign{\smallskip}
\hline
  \noalign{\smallskip}
 run ID & $\phi$ & $\psi$ & $\chi$ & $t_{\rm F}$ & Comment\\ 
   \noalign{\smallskip}
 \hline
 \noalign{\smallskip} 
 Scalar & 1.0  & 1.0 & 1.0 & 30 & as \citet{2014ApJ...796...29S}\\
 phi\_A & 1.0  & 1.0 & 2.0 & 10 & strong amplification \\
 phi\_B & 1.0  & 1.0 & 0.5 & 10 & weak amplification \\
 phi\_C & 1.0  & 1.0 & 0.1 & 10 & very weak amplification \\
 R\_A   & 2.0  & 1.0 & 1.0 & 10 & magnetic loops at $R\simeq40$ \\
 R\_B   & 0.75 & 1.0 & 1.0 & 10 & magnetic loops at $R\simeq20$\\ 
 th\_A  & 1.0  & 5.0 & 1.0 & 10 & multiple loops in $R\in[15,80]$ \\
 th\_B  & 1.0  & 0.1 & 1.0 & 4  & magnetic loops at $R\simeq15$ \\
 \hline
\end{tabular}
\label{tab::an_cases}
\end{table}

The role of the diffusivity has been widely discussed in the literature  \citep{2007A&A...469..811Z, 2012ApJ...757...65S}.
Here we assume 
\begin{equation}
    \overline{\eta}_0 = \left( \frac{1}{2}, \frac{1}{2}, 1 \right) \eta_0,
\end{equation}
where $\eta_0 = 0.165$ recovers the reference values of \citet{2014ApJ...796...29S}.
In order to have a direct comparison with the simulations of \citet{2018ApJ...855..130F}, 
we set the dynamo tensor components as
\begin{equation}
    \overline{\alpha}_0 = (\phi,\psi,\chi)\alpha_0,
\end{equation}
with $\alpha_0=0.775$.
Setting $\psi=\phi=\chi=1$, we recover the reference simulation of \citet{2018ApJ...855..130F}\footnote{Note that
$\alpha_{\rm{ss}}$ as well as the dynamo tensor (now also considering sound speed) are now differently defined.
Thus, the coefficients $\alpha_0$ and $\eta_0$ are not defined in the same way.}.

The strength of the dynamo coefficients $(\phi,\psi,\chi)$ are summarized in Table~\ref{tab::an_cases}.
From this set of simulation runs, we will consider a sample of eight exemplary runs in order to 
disentangle the influence of the different components of the alpha tensor on the magnetic field structure and the 
disk and jet evolution.

%==============================================================================
%
%
%
%==============================================================================

\begin{figure*}
\centering
\includegraphics[width=0.24\textwidth]{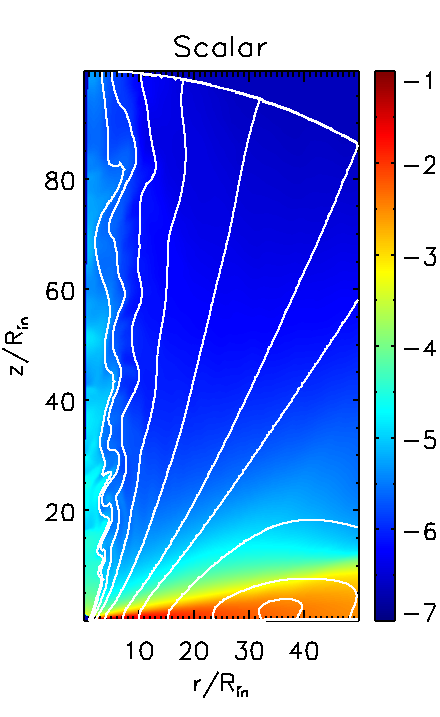}%
\includegraphics[width=0.24\textwidth]{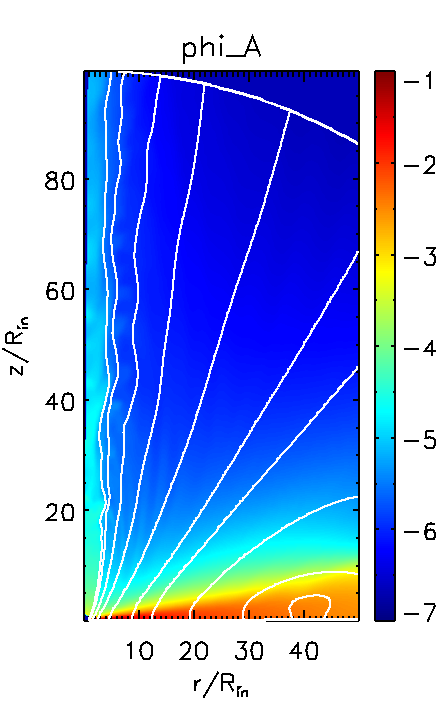}%
\includegraphics[width=0.24\textwidth]{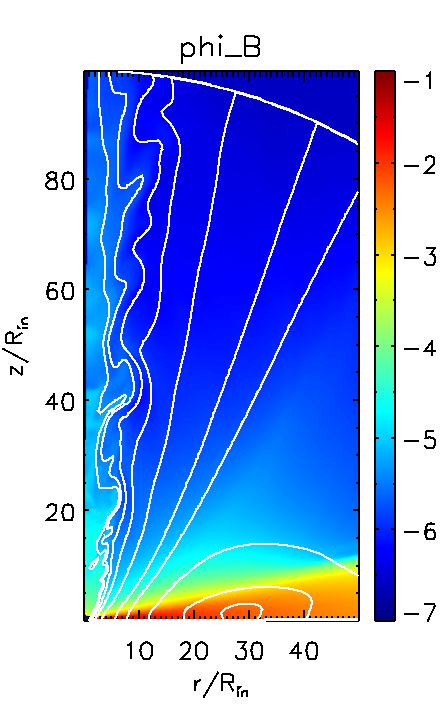}%
\includegraphics[width=0.24\textwidth]{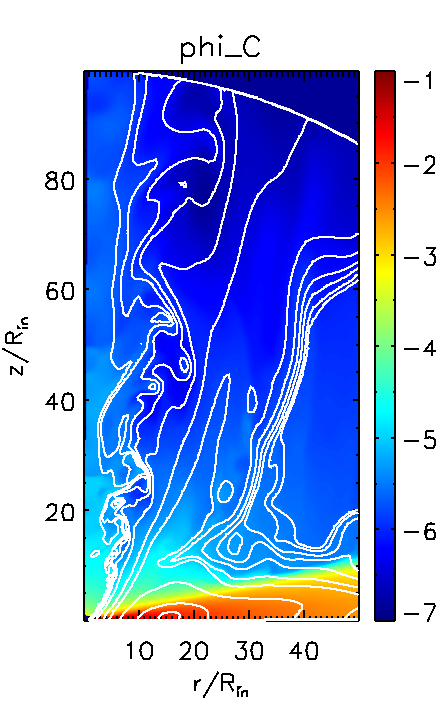}
\includegraphics[width=0.24\textwidth]{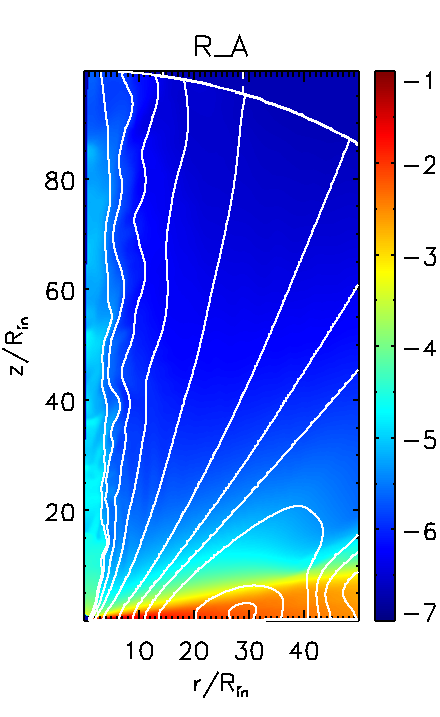}%
\includegraphics[width=0.24\textwidth]{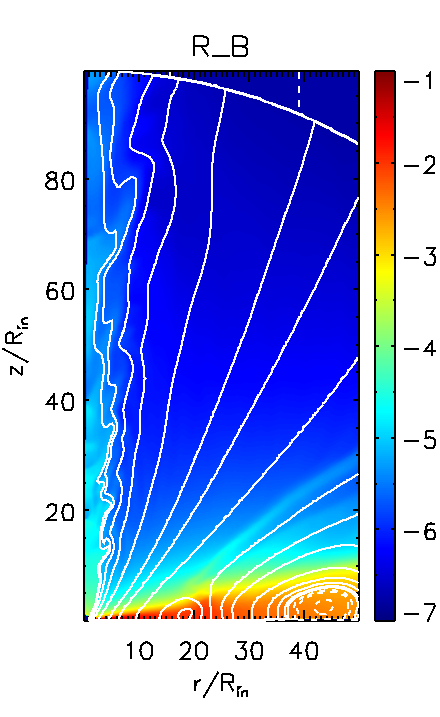}%
\includegraphics[width=0.24\textwidth]{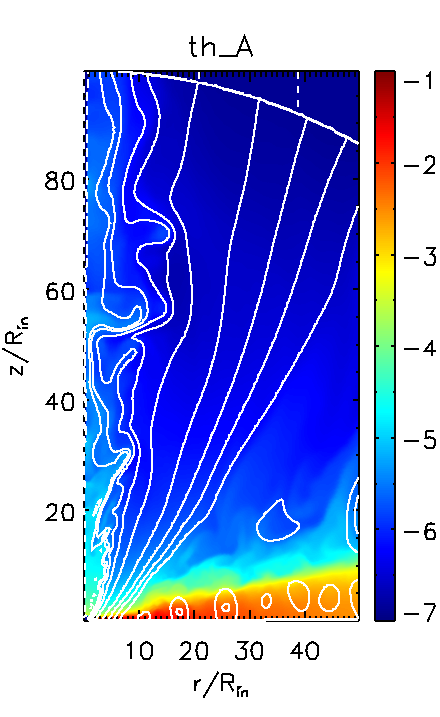}%
\includegraphics[width=0.24\textwidth]{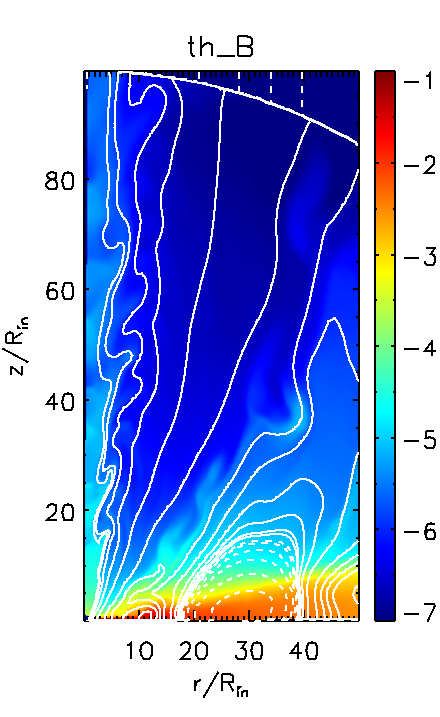}
\caption{ Magnetohydrodynamic evolution of the toy model dynamo simulations (see Table~\ref{tab::an_cases} at $t = 4000$.
Shown is the density distribution (color) and magnetic field lines (white lines).
The poloidal magnetic is field is represented by the contour lines of the vector potential $A_\phi$.
The dashed lines indicate a negative polarity of the poloidal magnetic field. }
\label{fig::rho_pol_toy}
\end{figure*}

%----------------------------------------------------------------------------------
\subsection{Evolution of the magnetic field}
Figure~\ref{fig::rho_pol_toy} shows for the different parameter runs the density distribution of the disk-jet structure, 
together with the magnetic field geometry (as contour lines of the vector potential).
We point out that in all cases but simulation {\em th\_B} (which is described more in detail in Section \ref{sec::toy_early}),
the initial magnetic field has the radial structure as described in Section \ref{sec::init}. 

Overall, we see that in all simulation runs the magnetic field in the inner region close to the rotation
axis that has been generated by dynamo action shows a large scale open geometry.
Together with a substantial strength, this magnetic field structure is able to eject disk material in to an 
outflow with a high degree of collimation.
On the other hand, we also see that the very field structure depends on the choice of the dynamo tensor, thus the 
strength of the tensor components.
The choice of different coefficients $(\phi,\psi,\chi)$ in our toy dynamo model leads to a different magnetic field
configuration.

%---------------------------------------------------------------------------------------

\subsubsection{A super-critical poloidal dynamo}
\label{sec::toy_crit}
The induction equation tells us that the dynamo action governed by $\alpha_\phi$ is the only way to increase
the poloidal magnetic field up to the strength that is required for jet launching.
Even for $\alpha_\phi = 0$ the toroidal magnetic field is still dynamo-amplified through the $\Omega$ effect 
and also the $\alpha_R$ dynamo component.
However, the dynamo does not lead to a substantial amplification of the poloidal magnetic field.
Therefore the latter cannot increase and stays confined within the disk.
Neither the strength nor the launching angle can be reached that is required to produce a Blandford-Payne 
outflow.

On the other hand, as a consequence of the quenching model applied, the magnetic diffusivity still increases as the
toroidal magnetic field growths. 
As a consequence, the poloidal magnetic field still evolves, even if the field is not enhanced by the dynamo 
action.
Note, however, that even if $\chi > 0$, if $\alpha_\phi$ is under a critical strength $\alpha_{\rm{crit}}$, 
the dynamo action for the poloidal field is still negligible.

When comparing the time scales for diffusion and dynamo action for different strength for $\alpha_\phi$
(see \citealt{2018ApJ...855..130F}),
we find a critical value of $\alpha_{\rm{crit}} \simeq 0.003$, corresponding to $\chi\simeq0.03$.

Because of the diffusivity model applied in these simulations, 
we  find that the dynamo number is not an unambiguous measure for the initial
critical dynamo action, as the disk diffusivity does not only depend 
on the {\em poloidal} magnetic field (that is not amplified as the 
dynamo $\alpha_\phi$ is sub-critical), but also on the {\em toroidal} 
magnetic field (that remains amplified by the dynamo $\alpha_R$ and by 
the $\Omega$ effect).
Moreover, as shown in \citet{1988ApJ...331..416S,1990ApJ...362..318S,1994A&A...283..677T},
the initial critical dynamo number depends on several factors e.g. the number 
of grid cells or the magnetic field configuration.
Nevertheless, the dynamo number still remains a key parameter in order to understand the evolution and
saturation of the dynamo action (see Sect.~\ref{sec::eq_dynnum}).

\begin{figure}
\centering
\includegraphics[width=0.45\textwidth]{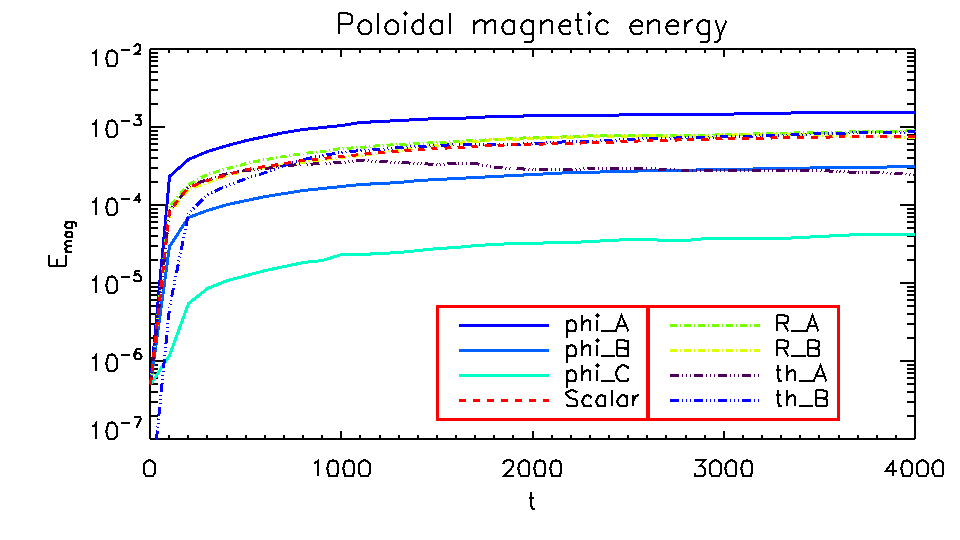}
\caption{Temporal evolution of the disk poloidal magnetic energy integrated
from radius $R = 10$ to the end of the domain, $R_{\rm{out}} = 100$.}
\label{fig::toy_poldisk}
\end{figure}

For $\alpha_\phi <\alpha_{\rm{crit}}$, the poloidal magnetic field increases only by less than one order of magnitude 
in the outer disk before time $t = 3000$, while the inner disk region is not at all magnetically amplified.
For $\alpha_\phi > \alpha_{\rm{crit}}$, the dynamo effect substantially amplifies the poloidal magnetic field, 
as shown in Fig.~\ref{fig::toy_poldisk}, 
changing both its strength and geometry which subsequently may lead to an disk outflow of material similar to 
\citet{2014ApJ...796...29S}.
All the cases that we investigated and that are listed in Table~\ref{tab::an_cases}
satisfy the condition $\alpha_\phi > \alpha_{\rm{crit}}$.

%==============================================================================
%
%
%
%==============================================================================

\subsubsection{Induction of multiple magnetic loops}
\label{sec::toy_loops}
Dynamo action triggered by the $\phi$-component of the alpha tensor leads to a topological magnetic field structure
such that the magnetic field loops generated in the inner disk region open up and drive a collimated outflow 
(see also Fig.~\ref{fig::rho_pol_toy}, but also \citealt{2014ApJ...796...29S, 2018ApJ...855..130F}).
Outside this inner jet launching region, magnetic loops are continuously formed.
This loop structure, that is basically corresponding to a reversal in the radial field $B_R$,
is diffusing outwards due to the radial magnetic field pressure gradient of the inner disk.
Such loops do not correspond to a reversal in the toroidal field and since they are diffused away, their impact on the jet dynamics is negligible.

In case of dynamo action that is substantially an-isotropic (as our cases {\em R\_A, R\_B} and {\em th\_A}), 
we observe an essentially different evolution of the magnetic field topology.
That is, for $\phi < 0.8$ or $\phi > 1.5$, {\em a second magnetic loop} is formed that is {\em anti-aligned} to the loop structure 
induced further in.
These loops, characterized by a reversal in the toroidal field, are substantially different from the ones described previously, 
and play a significant role in the evolution of the magnetic field and of the disk-jet system (see our discussion below).
We point out that the anti-aligned magnetic loops can be formed also when considering a scalar dynamo tensor, when 
the scalar $\alpha_0 < 0.6$ \citep{2018ApJ...855..130F}.

As shown in Section \ref{sec::toy_crit}, the coupling between the toroidal magnetic field and the dynamo tensor component
$\alpha_\phi$ is the main mechanism responsible for generation of the poloidal field.
For $0.8 < \phi < 1.5$, the toroidal magnetic field, being amplified by the $\Omega$-effect from the radial weak seed 
field, shows a monotonous behavior (after being amplified).
As the system evolves, the poloidal field is amplified over the whole accretion disk.

By looking at the spatial and temporal numerical derivatives of the toroidal field, we find that because of the highly 
anisotropic character of $\alpha_R$, some {"}dynamo-inefficient zones{"} are formed.
These are areas of vanishing poloidal field strength, but, in addition, in such zones also the toroidal magnetic field cannot be amplified.
The number and the location of these zones, where the dynamo is not efficient, depends on the strength of the three dynamo components and not exclusively by $\alpha_R$.

\begin{figure*}
\centering
\includegraphics[width=0.33\textwidth]{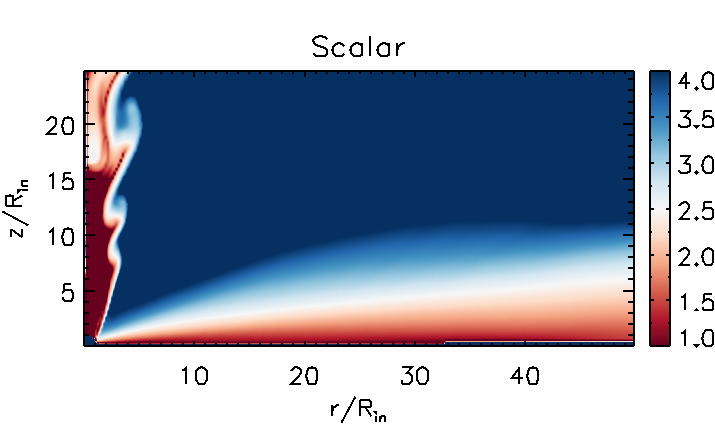}
\includegraphics[width=0.32\textwidth]{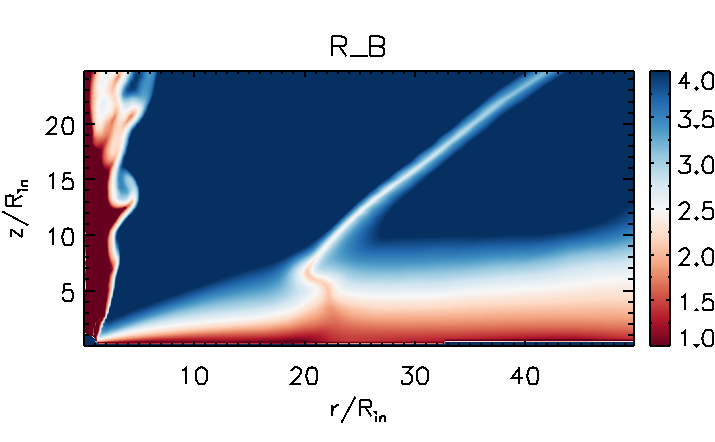}
\includegraphics[width=0.33\textwidth]{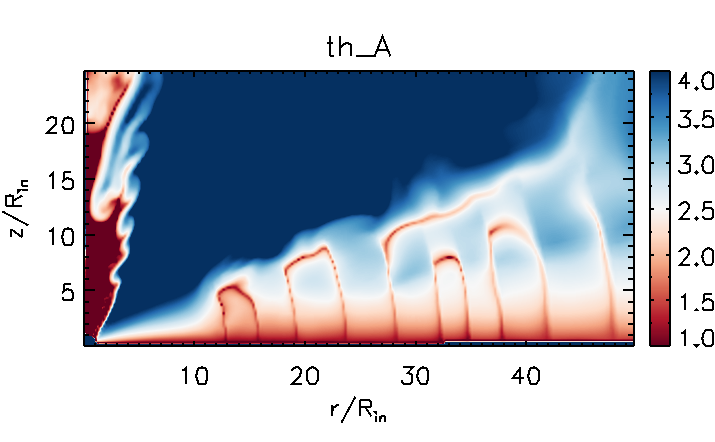}
\includegraphics[width=0.32\textwidth]{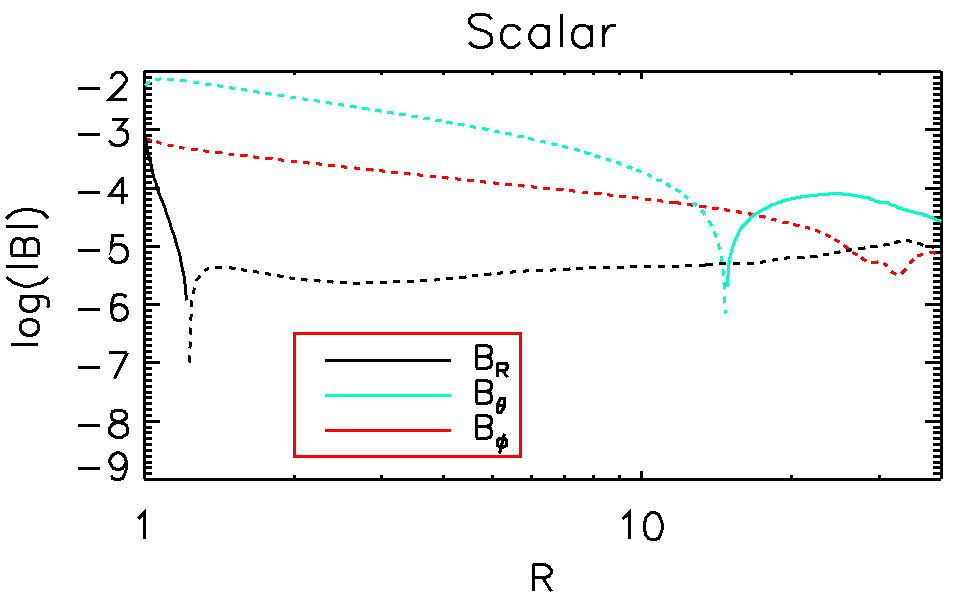}
\includegraphics[width=0.33\textwidth]{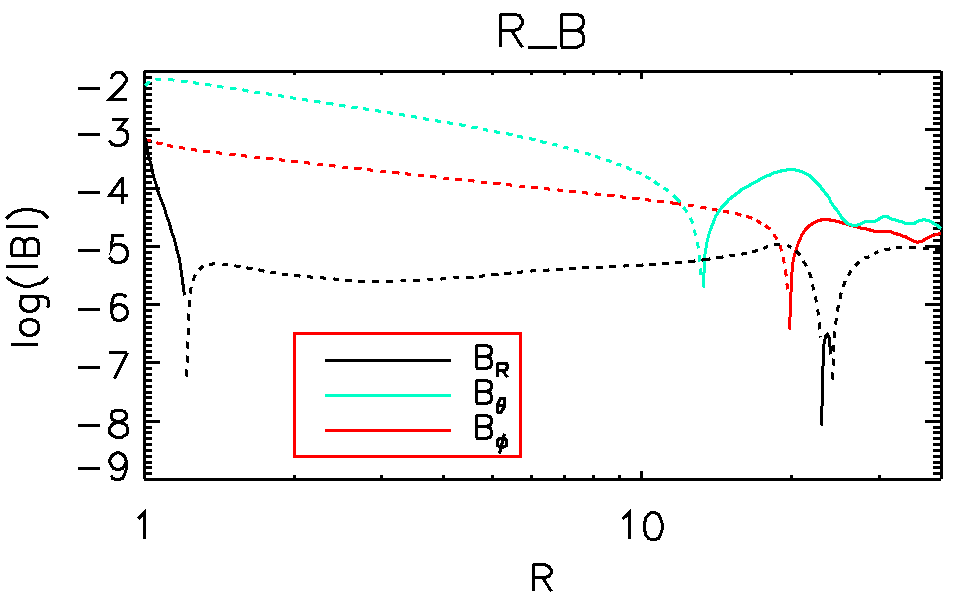}
\includegraphics[width=0.32\textwidth]{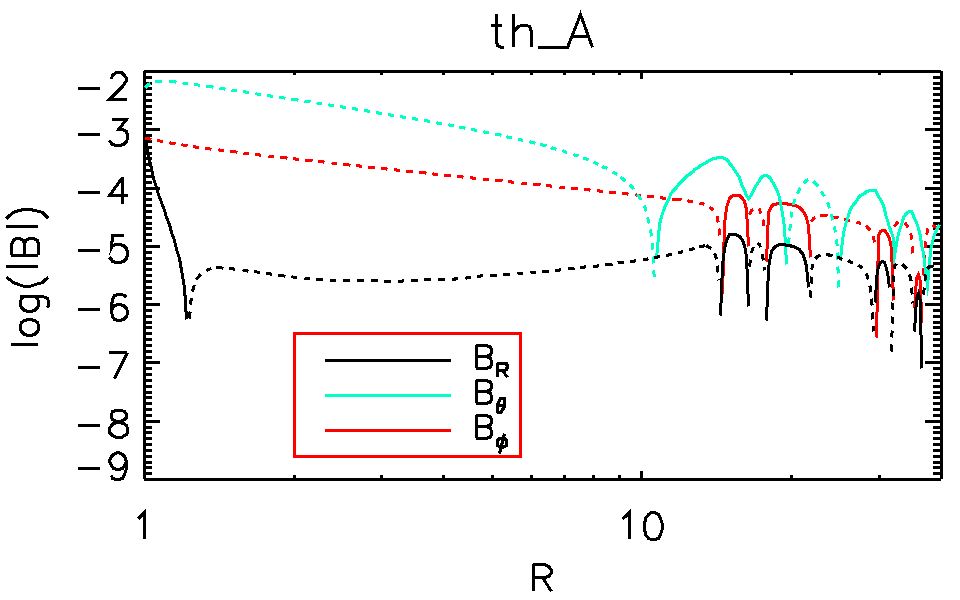}
\caption{Presence and absence of dynamo-inefficient zones in the disk for different dynamo prescriptions, 
{\em Scalar} (left panels, {\em R\_B} (middle panels), and {\em th\_A} (right panels), respectively. 
The top row shows the disk magnetization at time $t = 4000$, 
while in the bottom row shows the three magnetic field components 
close to the disk mid-plane at time $t = 1000$, 
where the solid lines represent positive values of the magnetic field and the dashed lines represent negative values 
of the magnetic field.}
\label{fig::toy_loops_disk}
\end{figure*}

Furthermore, for $\alpha_\theta > 3$, we find that the toroidal field shows multiple dynamo-inefficient zones.
On the other hand, the dynamo-inefficient zones of case {\em th\_A} remain confined in the accretion disk.
This is illustrated in the top panels of Fig.~\ref{fig::toy_loops_disk} where we show the disk magnetization at the same 
evolutionary time, $t = 4000$.
The difference between between the three simulation runs {\em Sc}, {\em R\_B}, and {\em th\_A} is clearly visible.

For the case of the scalar dynamo the local disk magnetization is only weakly dependent on the radius.
It is relatively low along the mid-plane and increases towards the disk surface.
This is understandable as the disk gas pressure decreases with altitude while the poloidal field remains rather constant vertically.

For simulation run {\em R\_B}, for which $\alpha_R = 0.75$, we see that a dynamo-inefficient zone has developed
around radius $R\simeq23$.
Typically, these zones seem to be anchored at the disk mid-plane.
As they are balanced by a low magnetic pressure, they vertically extend while preserving the total pressure equilibrium.

For simulation run {\em th\_A}, for which  $\phi = 1$ and $\psi = 5$, we find multiple dynamo-inefficient zones along 
the accretion disk.
Note that due to their proximity, these zones are able to connect -- and reconnect. 

Because the coupling between the toroidal field and the $\alpha_\phi$-component of the dynamo tensor is the only way
to dynamo-amplify the radial field component, the radial field that is amplified from the toroidal has also
different polarities.

Since we have physical resistivity included, the magnetic field is able to reconnect and to change its topology within 
the accretion disk.
In particular, instead of one magnetic loop that is visible
(see Fig.~\ref{fig::toy_loops_disk}, bottom right panel),
now more magnetic loops arise (see Fig.~\ref{fig::toy_loops_disk}, bottom left panel).
On the other hand, the reversal of the toroidal field is associated with a maximum in the tensor component 
$\alpha_\theta$, 
which undergoes a reversal at smaller radii (bottom left panel of Fig.~\ref{fig::toy_loops_disk}).

Compared to the results of \citet{2018ApJ...855..130F}, here we find that the re-configuration of the magnetic field 
structure does impact the jet evolution on a weaker level.
We believe that this is mainly due to the mid-plane boundary condition, that is absent in the previous paper.
In particular, here we enforce symmetry between upper and lower hemisphere that can be violated in a bipolar setup.
However, the reversal of the toroidal and radial field components that directly define the disk magnetization
still play a key role in the disk-jet evolution.
We note that the disk magnetization is the main ingredient of the diffusivity model for the resistive disk evolution.

Since the dynamo-inefficient zones correspond to zones of low diffusivity, as a result the accretion process can be 
affected.
In fact, accretion can be suppressed across such zones, leading to under-dense and over-dense regions (compared to 
the simulations without multiple loops). 
We find that these under-dense/over-dense regions are strongly related to the existence of a vertical field.
We experienced numerical issues when under-dense zones are located too close to the inner boundary, for example
unphysical values of the fluid density or the fluid pressure.

Here we need to comment briefly on the {"}dead zones{"} that has been proposed for protoplanetary disks.
Although the dynamo-inefficient zones we detect in our simulations may look similar to these dead zones,
the physical processes involved are not the same.
Dead zones in protoplanetary disks have been proposed by \citet{1996ApJ...457..355G} on the basis of a lack
of coupling between matter and magnetic field due to an insufficient degree of ionization.
This lack of coupling would not allow the MRI to operate, and, as a consequence, also accretion
unlikely to happen, since the lack of angular momentum exchange. 
As a result, a layered accretion is expected on a theoretical basis, which could indeed be realized in numerical simulations
\citep{2000ApJ...530..464F,2003ApJ...585..908F}.
Also, resistivity was found to play an essential role in suppressing the MRI 
(see e.g. \citealt{2000ApJ...543..486S,2002MNRAS.329...18F, 2012ApJ...761...95F}).
Dead zones in protoplanetary disk are also thought to be responsible to create
transition disks \citep{2016A&A...596A..81P}.

It is interesting to note that for both the protoplanetary dead zones and for our dynamo-inefficient zones
the resistivity plays a leading role.
However, for the first approach it is the resistive de-coupling which suppresses the MRI (and would subsequently 
suppress the dynamo action of the MRI), while for our models the dynamo-inefficient zone is formed as result of
a minimum of the magnetic diffusivity.

Finally we note that as the dynamo-efficient zones are basically resulting from the feedback of the magnetic field on the magnetic diffusivity, a change in the quenching model -- from the diffusive quenching to the standard quenching -- may 
affect the exact location and width of the dynamo-inefficient zones.

%-----------------------------------------------------------------------------------
\subsubsection{Amplification of the magnetic field}
\label{sec::toy_bmag}
The majority of our parameter runs apply a super critical dynamo $\alpha_\phi > \alpha_{\rm{crit}}$ 
(Table~\ref{tab::an_cases}).
The resulting magnetic field strength and geometry supports a collimated outflow.
In Fig.~\ref{fig::toy_poldisk} we show the time evolution for the disk poloidal magnetic energy, integrated from $R = 10$.
For a comparison the case of an isotropic dynamo is shown.

The three different dynamo tensor components play a different role in the amplification of the poloidal magnetic field.
The $\phi$-component of the dynamo is the main ingredient that amplifies the poloidal magnetic field in 
the disk, while the $R$ and $\theta$-components determine the formation of the dynamo-inefficient zones, 
that, subsequently, also determines the poloidal magnetic field structure.

The $\phi-$component of the dynamo tensor essentially influences already the very early stages of the 
disk-jet evolution -- a higher strength of $\alpha_\phi$ leads to a faster and stronger amplification,
as we can see by comparing the {"}{\em phi}{"}-simulations to the isotropic model in Fig.~\ref{fig::toy_poldisk}.

The other dynamo components ($\alpha_R$ and $\alpha_\theta$) become important only once the poloidal field has been 
amplified to substantial strength, and through the presence (or absence) of the dynamo-inefficient zones.
In particular, where a dynamo-inefficient zone is built up in the {\em inner} disk, it triggers the temporal evolution of the system already on short timescales ($\simeq100$ after its formation ).
A dynamo-inefficient zone located further out plays a minor role during the early phase of the disk evolution.

We emphasize that the evolution of the disk magnetic field is strictly 
correlated with the existence of dynamo-inefficient zones, since these features
lead to the formation of multiple anti-aligned magnetic loops in the disk (see Fig.~\ref{fig::rho_pol_toy} and section \ref{sec::toy_loops}).
A higher strength of the dynamo tensor component $\alpha_R$ leads to an -- on average -- higher amplification of the 
toroidal field.
However, once the dynamo is quenched by magnetic diffusivity, the magnetic field strength decreases to the magnitude
that we recovered in the isotropic dynamo simulation.
Therefore, we interpret that the effect of a higher $\alpha_R$ is a more rapid amplification of the poloidal field.
On the other hand, a lower $\alpha_R$ leads to a slower toroidal (and therefore poloidal) field amplification.

We find a different behavior when a dynamo-inefficient zone (only one) is forming which extends beyond the accretion disk surface.
As discussed in Section \ref{sec::toy_loops}, the reversal of the toroidal field corresponds to a spatially stationary point 
in the $\theta-$component of the magnetic field.
As a result, the poloidal magnetic energy is higher than for the isotropic dynamo model, simply because in the 
dynamo-inefficient regions of the disk the vertical field component becomes stronger.

On the other hand, this increase in the vertical component of the magnetic field is partially suppressed in the 
presence of multiple magnetic zones, 
compared to the case of an isotropic dynamo tensor.
Our understanding of this effect is that the existence of quite a number of field reversals 
(that effectively decrease of the local magnetic energy), 
more than compensates the induction of a vertical field component (that would lead to a decrease of
the local magnetic energy).

%---------------------------------------------------------------------------------
\subsubsection{The dynamo number}

\begin{figure*}
\centering
\includegraphics[width=0.24\textwidth]{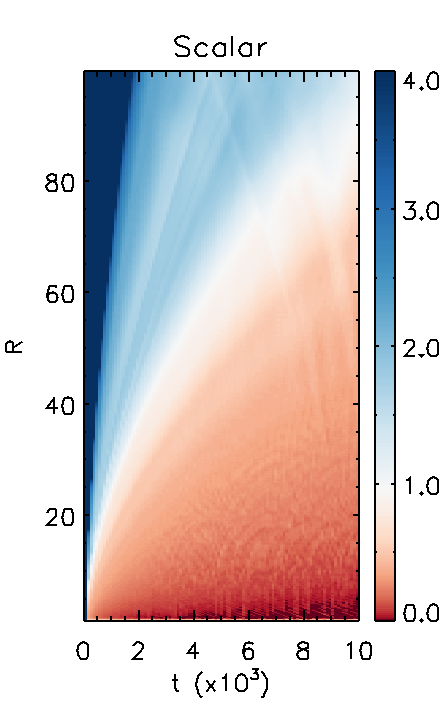}
\includegraphics[width=0.24\textwidth]{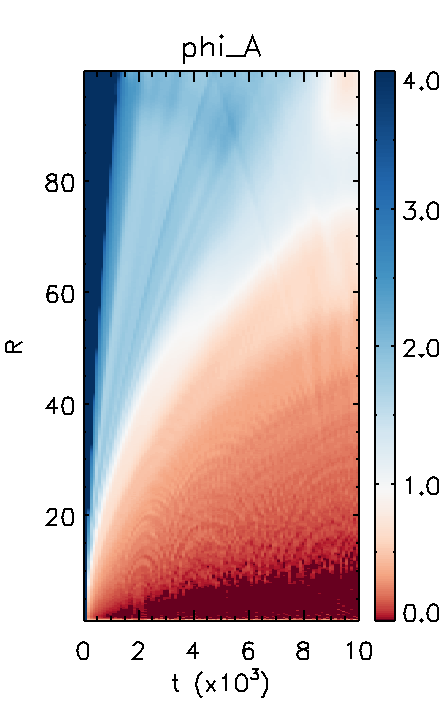}
\includegraphics[width=0.24\textwidth]{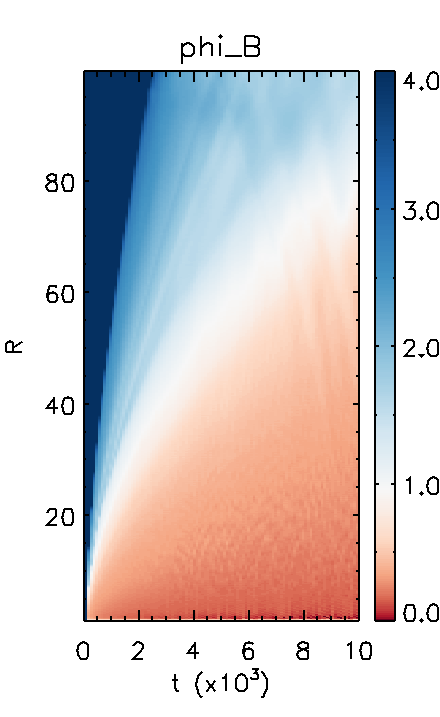}
\includegraphics[width=0.24\textwidth]{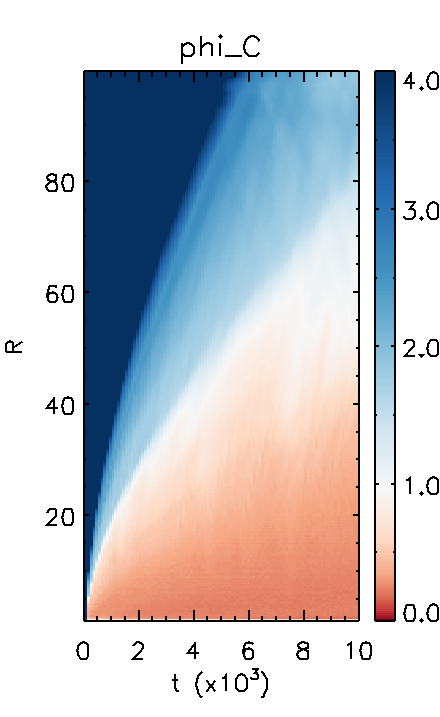}
\includegraphics[width=0.24\textwidth]{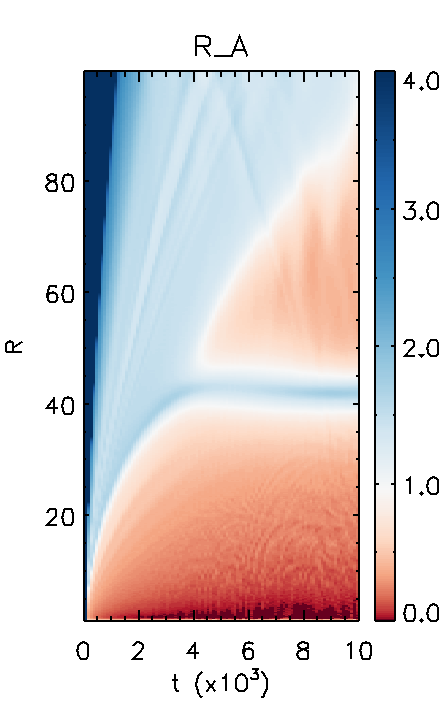}
\includegraphics[width=0.24\textwidth]{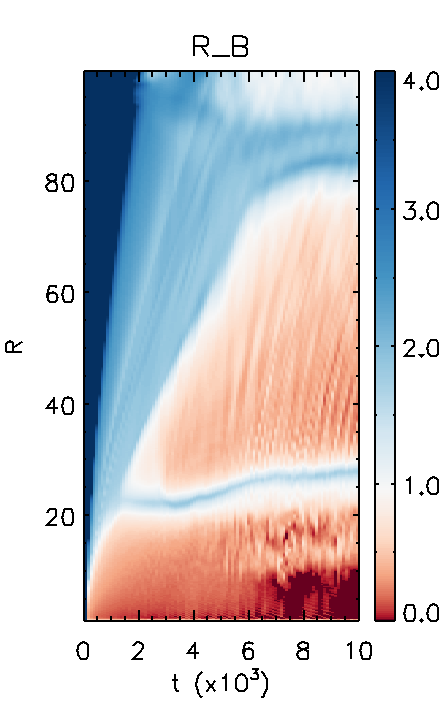}
\includegraphics[width=0.24\textwidth]{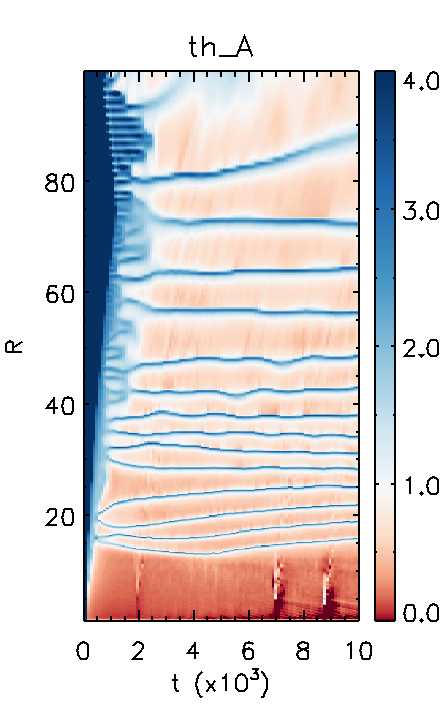}
\includegraphics[width=0.24\textwidth]{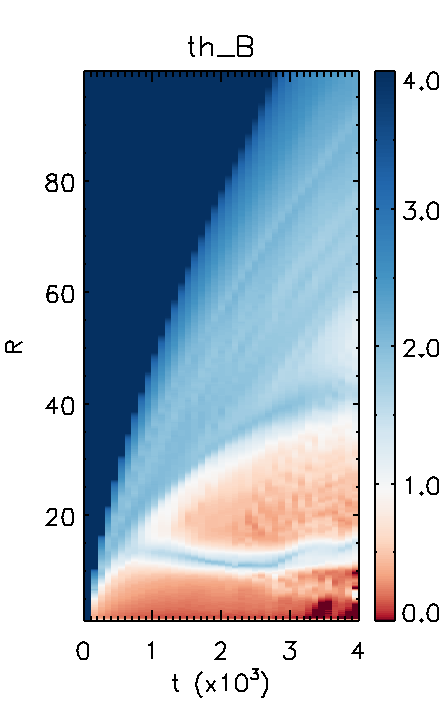}
\caption{Dynamo number for selected simulation runs.
Note the presentation of this figure as a t-R diagram, displaying the strength dynamo number along the disk 
(vertically, in R-direction) as a function of time (horizontal axis).}
\label{fig::toy_dynnum}
\end{figure*}

The dynamo number is usually quoted as a measure for dynamo activity.
Only dynamos with a super-critical dynamo number evolve rapidly and work efficiently against magnetic diffusivity, 
and finally lead to a strong, saturated poloidal magnetic field.
The dynamo number can therefore tell us when and where the growth of the 
magnetic field reaches saturation.
In Figure~\ref{fig::toy_dynnum} we compare the dynamo number as function of time and radius for different cases.

We first show the dynamo number for different strength of the tensor component $\alpha_\phi$ (top panel).
We see that as $\chi$ decreases, the amplification of the poloidal field is weaker and also slower, as also indicated
by Fig.~\ref{fig::toy_poldisk}.
These differences in the magnetic field evolution are reflected on the dynamo number.
In the time evolution of the dynamo number for all simulations we can clearly distinguish three evolutionary
stages\footnote{Here, we point out that, as opposed to the simulations described in the lower panel of
Fig.~\ref{fig::toy_dynnum}, the top panel is marked by the absence of the multiple magnetic loops described 
in Section \ref{sec::toy_loops}.
For this reason the three evolutionary stages we prefer to define considering both time and space.}.

We may first define an (i) initial phase (indicated in blue) during which the dynamo number is almost infinite, simply
because the diffusivity is still low (as implied by the quenching triggered by the magnetic diffusivity).
Then comes a  (ii) dynamo phase (indicated in white) that is characterized by a strong competition between dynamo
action and diffusive quenching.
During this phase we recognize magnetic loops being present, surviving from the early stages
($t\lesssim500$ in the inner disk region) of dynamo evolution.
In a subsequent (iii) final phase (indicated in red), these magnetic loops have been washed out or have been
broken-up, respectively, and a quasi-steady state of the dynamo evolution is reached.
The time scale when the final phase is reached depends of the radius (thus on the dynamical time scale
that is defined by the disk rotation at this radius).
In the inner radii the final stage is reached around $t\lesssim500$, 
while in the outer disk regions is reached only at  $t\gtrsim5000$.
In this final phase, dynamo action and diffusive quenching are fully balanced.

Note that in the inner disk region the second dynamo phase is missing because of the rapid evolution of the dynamo.
Here, the magnetic energy reaches the saturation level already very early, with a timescale of the first two phases 
being much smaller.

Considering now the effect of different levels of dynamo an-isotropy we find the following results.
For larger $\chi$ we do not find a second phase at larger radii since the magnetic field is amplified on a shorter
timescale.
In addition to that, for larger $\chi$ the first phase has a shorter lifetime at every radius.

Looking at the innermost parts of the accretion disk, a larger $\chi$ leads to an overall smaller dynamo number
at the stage of quasi-steady state. 
This is a consequence of the quadratic dependence on the disk diffusivity (see Eq.~\ref{eq::ssm}) that balances, 
respectively quenches the mean-field dynamo effect.
For the latest evolutionary stages we notice that, although this happens at different times, 
for each choice of $\chi$,the simulation reaches its steady stage also at a larger radius.
This is an indicator of a faster evolution of the magnetic field for larger $\chi$.

Note that the dynamo number can also be used as a tracer to identify the dynamo-inefficient zones.
As the latter correspond to a minimum in the magnetic diffusivity, here the dynamo number will have a sudden growth.
On the other hand, the dynamo-inefficient zones are not only zones where the toroidal magnetic field has a minimum, 
but they also zones where the toroidal field cannot be amplified.
For such reason, the general application of the dynamo number as a measure of dynamo activity can be misleading,
since its sudden growth (in correspondence of the field reversal) does not necessarily lead to a further magnetic 
field amplification.

This is shown in Fig.~\ref{fig::toy_dynnum}, where we display in the bottom panels the dynamo number for the 
simulation runs
that result in the generation of dynamo-inefficient zones.
In contrast, the upper panels show simulations that do not lead to dynamo-inefficient zones.
The figure nicely demonstrates a similar evolution of these simulation up to radii where the dynamo-inefficient
zones have established when a quasi-steady state is reached.

Interestingly, the dynamo-inefficient zones -- representing a minimum in the toroidal and in the radial magnetic 
field component, do not directly affect the dynamo activity further out.
Outside the field reversal zone a saturation of the magnetic field can be reached.
This is in particular visible when comparing the two right panels (runs {\em phi\_C} and {\em th\_B}).

Looking at the dynamo number in more detail, we understand why case {\em R\_A} and case {\em Scalar} (isotropic dynamo) 
are almost not distinguishable (Fig.~\ref{fig::toy_poldisk}, left).
The dynamo-inefficient zone that is present in the case {\em R\_A} appears only at later stages of the evolution, 
as it is located at about $R\simeq40$ while the magnetic field in the ambient parts of the disk is amplified only 
on a longer timescale.
In contrary, the dynamo-inefficient zone of {\em R\_B} is formed already earlier at $t\simeq1000$, and therefore a 
different evolution of the poloidal disk magnetic field takes place, and also on a shorter timescale.
The time evolution of cases {\em th\_A} and {\em th\_B} will be discussed below (see Section \ref{sec::toy_early}).

%==============================================================================
%
%
%
%==============================================================================

\subsection{Dynamics of accretion-ejection}
So far we have investigated mainly the evolution of the magnetic field structure that is generated by the
accretion disk dynamo, applying different model assumptions for the dynamo tensor.
Obviously, the difference in the field structure - difference in strength and geometry - will have strong impact on the 
dynamics of the accretion disk and the disk wind or jet.
In this section we want to discuss the dynamical evolution of the accretion-ejection structure and 
compare the results for different dynamo models.

%------------------------------------------------------------------------
\subsubsection{Accretion and ejection rate}
\label{sec::toy_accrej}
\begin{figure}
\centering
\includegraphics[width=0.48\textwidth]{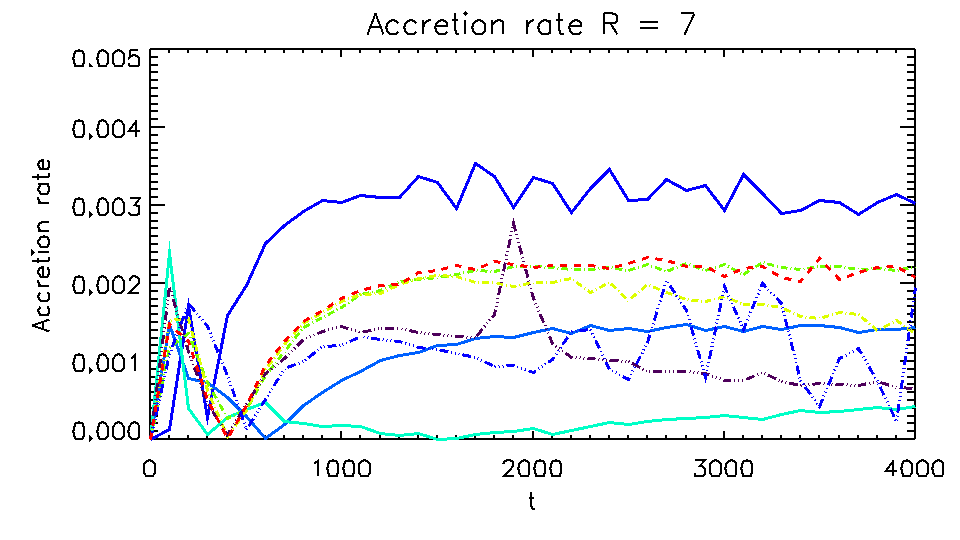}
\includegraphics[width=0.48\textwidth]{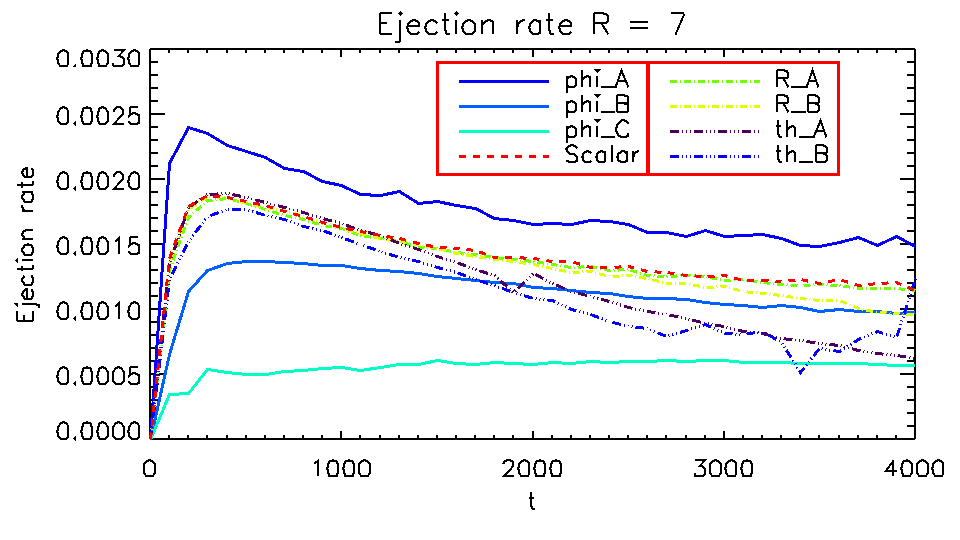}
\caption{Temporal evolution of the accretion (top panel) and ejection (bottom panel) rates. 
The accretion rate is computed at fixed radius $R = 7$, while the ejection rate is computed along the disk surface 
from $R = 1$ to $R = 7$.
See Sect.~\ref{sec::integrals} for a definition of the control volume.}
\label{fig::toy_accr_ej}
\end{figure}

As pointed in the previous sections, the dynamo tensor components that amplify the toroidal field ($\alpha_R$ 
and $\alpha_\theta$) 
work on longer timescales than the $\phi$-component of the dynamo tensor (which amplifies the poloidal magnetic field).
Also, a larger dynamo component $\alpha_\phi$ leads to a higher magnetic diffusivity.
In turn, this leads to a higher accretion rate, as shown in the top panel of Fig.~\ref{fig::toy_accr_ej}, since the 
disk diffusivity enables to replenish the disk matter that is 
lost from the inner disk (by accretion or ejection) from the outer disk regions.

On the other hand, the ejection rate only weakly depends on the strength of the $\phi$-dynamo, especially in the 
early stages of the evolution, ($t\simeq100$), as shown in the bottom panel of Fig.~\ref{fig::toy_accr_ej}.
While the inner regions reach a quasi-steady state for $t\gtrsim500$, the ejection rate decreases until it reaches
a quasi-constant level.
This magnitude is higher for larger $\chi$, mostly because of the enhanced accretion rate.

We find that the ratio between the ejection and the accretion rate is higher for lower $\chi$.
This can be understood as follows.
A higher strength of $\alpha_\phi$ leads effectively to a stronger and faster amplification of the magnetic field.
A larger $\chi$, which is itself a consequence of applying an anisotropic dynamo tensor, leads to a stronger disk 
magnetization\footnote{Note that the disk gas pressure, in absence of dynamo-inefficient zones, is subjected to 
only very small changes during the temporal evolution of the accretion disk.}.
Because of the diffusive quenching we apply (see Eq.~\ref{eq::ssm}), a higher disk magnetization implies a higher
disk magnetic diffusivity, which in turn supports higher accretion rates.

For example, Figure~\ref{fig::toy_accr_ej} shows that for $\chi=2.0$ about  $<50\%$ with of the accretion mass flux 
becomes ejected.
For $\chi<0.5$  all the matter accreted becomes ejected into the jet structure.
This result is in nice agreement with resistive non-dynamo launching simulations 
\citep{2007A&A...469..811Z, 2012ApJ...757...65S}, which showed a correlation between the disk magnetic diffusivity 
and the ejection-accretion rate ratio.

Once the poloidal field has become dynamo-amplified, the $R$ and $\theta$-components of the dynamo tensor can 
play a major role
in the magnetic field evolution and, thus, also in the dynamics of accretion-ejection as they potentially induce 
dynamo-inefficient zones in the disk.
For simulations for which NO dynamo-inefficient zones emerge, differences in the toroidal magnetic field do not really 
impact on the poloidal field components, even on longer time scales.

On the other hand we find that a toroidal field reversal and the subsequent formation of multiple anti-aligned 
loops (and the correspondent dynamo-inefficient zones) in the disk leads to a decrease in the accretion rate.
The reason is the diffusive quenching we apply.
At the locations where the toroidal field vanishes in the disk, also the magnetic diffusivity has a minimum (because of 
the low disk magnetization, see Eq.~\ref{eq::ssm}). 
A low diffusivity lowers the accretion efficiency.

We point out that the increase in the poloidal magnetic energy shown in Fig.~\ref{fig::toy_poldisk} is a value integrated
over a control volume.
Therefore, even if the overall magnetic energy is high, the formation of zones of low magnetic diffusivity leads to 
a decrease in the overall accretion rate.
As a consequence, the disk mass that is lost by accretion and ejection cannot be efficiently replenished, therefore the 
accretion rate decreases with time.
Also the ejection rate is affected, but at later times.
The most immediate consequence of the lower accretion rate is the formation of under-dense and over-dense zones in 
the accretion disk.

The radial distance of a dynamo-inefficient zone from the inner disk radius is strictly correlated with the timescale 
at which we observe a decrease in the accretion rate.
This is the case for example for simulations {\em R\_B} and {\em R\_A} (see Fig.~\ref{fig::toy_accr_ej}, left).
While in the former case the dynamo-inefficient zone leads to a decrease in the accretion rate already at time 
$t\simeq 2000$, 
the latter case shows no difference to the simulation applying an isotropic dynamo tensor until time $t\simeq4000$.
Note that the dynamo-inefficient zone is formed only at $t\simeq4000$, and, therefore, can impact the accretion 
and ejection rates only on a longer time scale (see Fig.~\ref{fig::toy_dynnum}).

%-------------------------------------------------------------------------
\subsubsection{Jet speed and collimation}
\label{sec::toy_jet}
An immediate consequence of a variation in the dynamo tensor components is the jet kinematics.
As pointed by \citet{2016ApJ...825...14S}, a higher poloidal disk magnetization will leads to a stronger jet, 
for example in terms of mass flux and velocity.
We know from simulations applying a scalar dynamo model \citep{2018ApJ...855..130F} that the terminal jet speed is 
correlated with the strength of $\alpha_0$; in particular, a stronger dynamo leads to a faster jet.
Note that these properties -- jet speed, mass flux, or collimation -- are global properties and thus accessible 
in principle by observations, different from the intrinsic local conditions in the disk such as turbulence and dynamo action.

\begin{figure}
\centering
\includegraphics[width=0.48\textwidth]{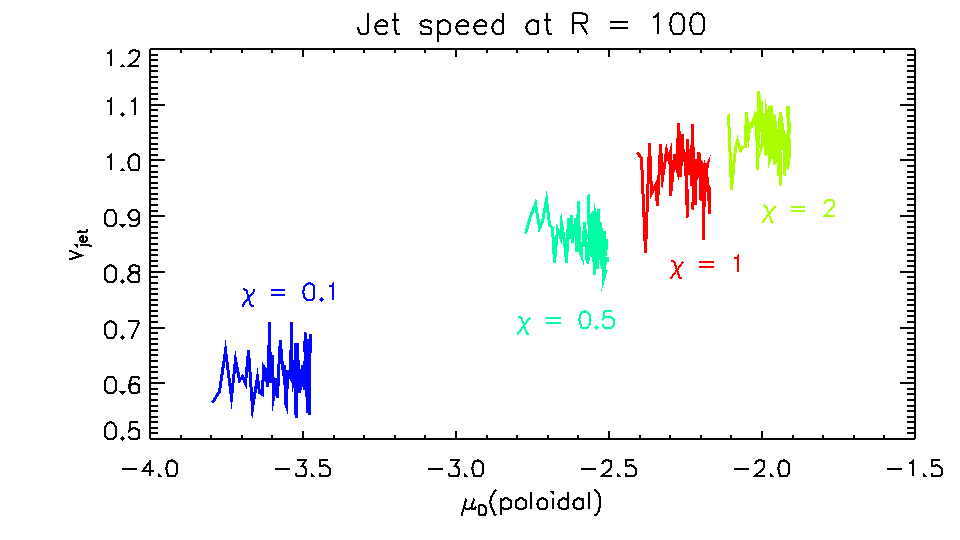}
\caption{Jet speed vs disk magnetization. 
Shown is the maximum jet velocity versus the disk magnetization calculated from the poloidal magnetic field.
Note that the disk magnetization is solely resolution from the dynamo component $xi$ and does not depend from a further quenching parameter.}
\label{fig::toy_velmag}
\end{figure}

As for the evolution of the magnetic field, the three components of the dynamo tensor have a different impact also
for the jet kinematics.
When considering different magnitudes of the dynamo-$\chi$, from our simulations we find an correlation similar
to the one discovered in \citet{2016ApJ...825...14S}.
That is the fact that a stronger $\phi-$component of the dynamo results in a stronger amplification of the 
poloidal magnetic field.
As a direct consequence, since the midplane pressure shows only a very weak dependence on the dynamo model, 
a larger $\chi$ leads to a higher poloidal disk magnetization (see Fig.~\ref{fig::toy_velmag}).
Consequently, with a higher disk magnetization more magnetic energy is available to accelerate the outflow.

\begin{figure*}
\centering
\includegraphics[width=0.23\textwidth]{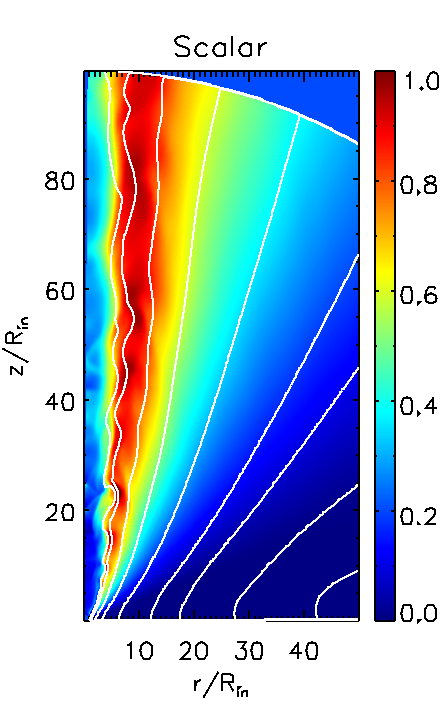}%
\includegraphics[width=0.23\textwidth]{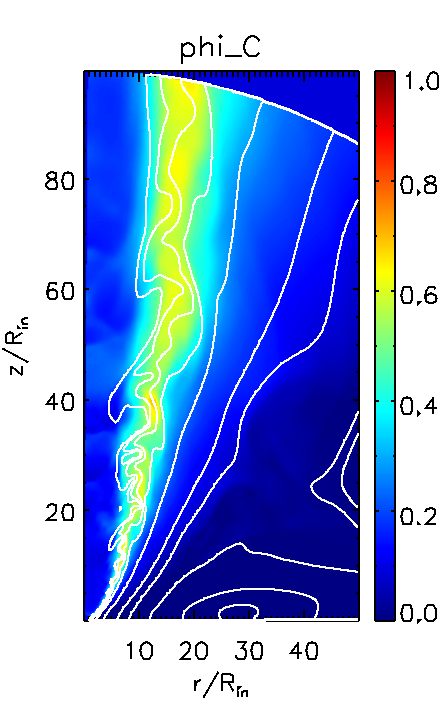}%
\includegraphics[width=0.23\textwidth]{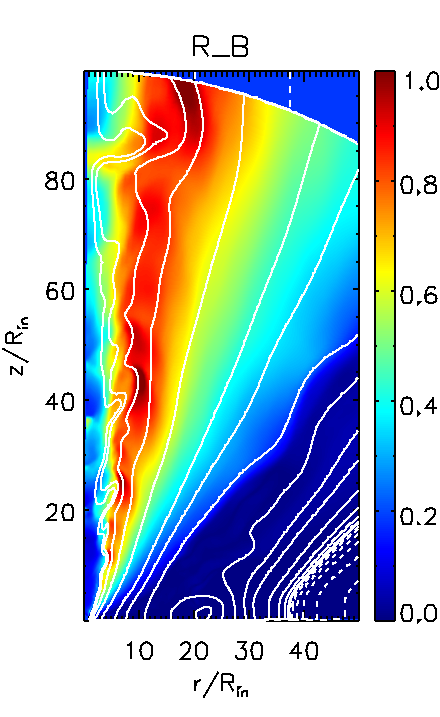}%
\includegraphics[width=0.23\textwidth]{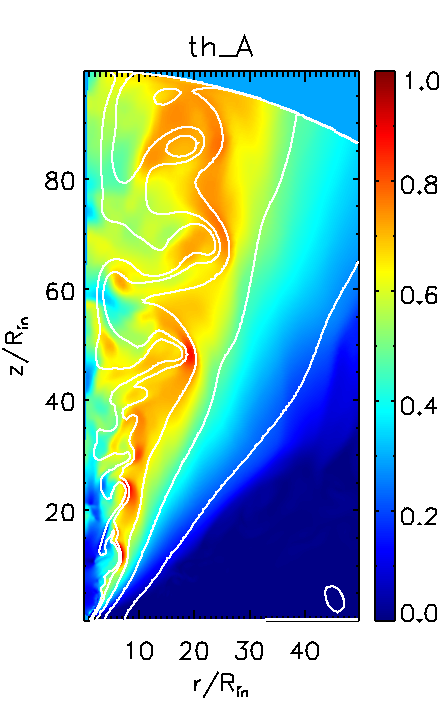}
\includegraphics[width=0.23\textwidth]{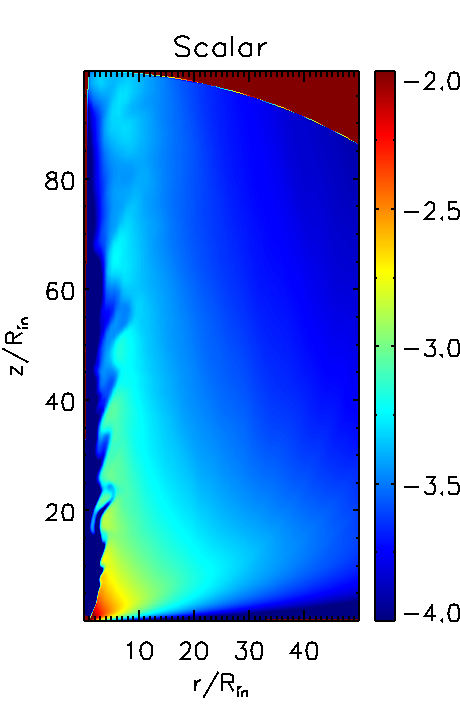}%
\includegraphics[width=0.23\textwidth]{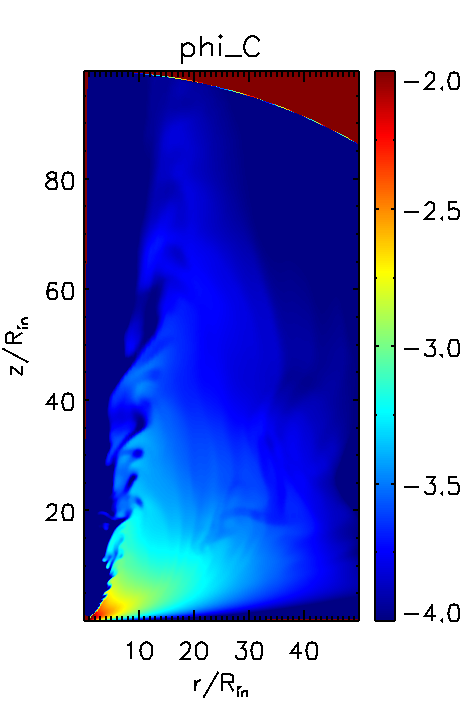}%
\includegraphics[width=0.23\textwidth]{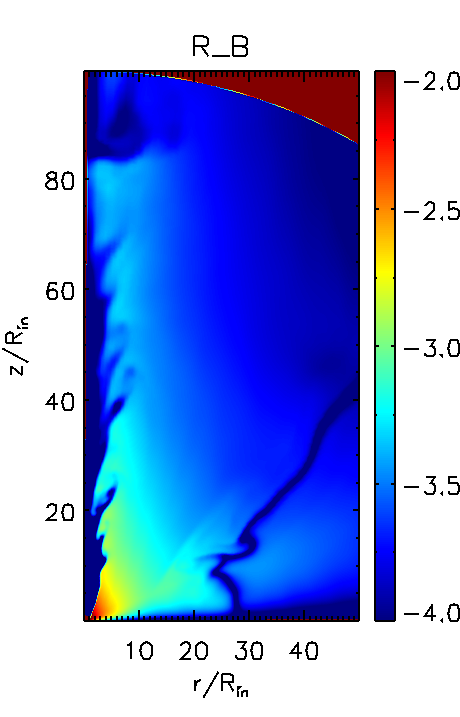}%
\includegraphics[width=0.23\textwidth]{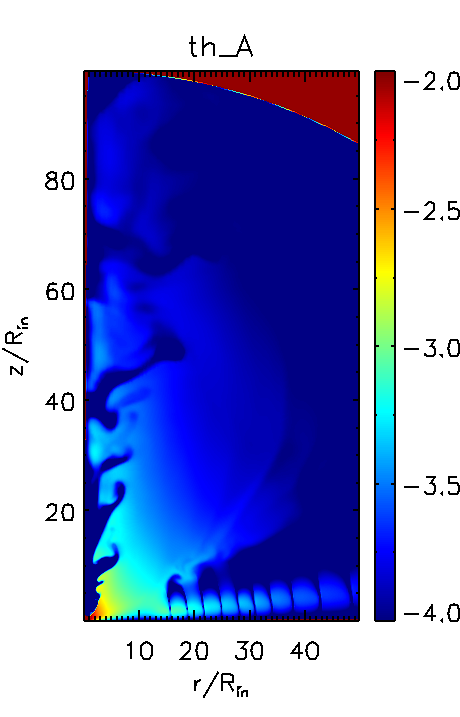}
\caption{Comparison of parameter runs at $t = 10000$.
 Shown is the distributions of the poloidal velocity (top), overlaid with contour lines of the vector 
 potential (following poloidal field lines)(top), and toroidal magnetic field strength (bottom).}
\label{fig::toy_jet_speed}
\end{figure*}

We show the terminal jet speed, here computed as the maximum speed at $R = 100$, as function of the 
magnetization in Fig.~\ref{fig::toy_velmag}.
This figure indicates a very clear trend, as proposed by \citet{2016ApJ...825...14S}.
In addition, it demonstrates again the gain in magnetization for 
different parameters for the dynamo parameter.
We find that the maximum jet speed reaches the Keplerian velocity at the inner disk radius
However, the maximum speed decreases for smaller $\chi$.
This is shown also in Fig.~\ref{fig::toy_jet_speed} where we compare the distribution of the jet poloidal velocity 
for different simulation runs.

The two other dynamo tensor components affect the evolution of the disk magnetization in term of generation 
(or not generating) magnetic loops and/or dynamo-inefficient zones.
Since minima in the magnetic field strength do only have a very minor impact on the overall disk poloidal magnetic 
energy (and therefore on the disk poloidal magnetization, see Fig.~\ref{fig::toy_accr_ej}), 
a difference in $\phi$ does not necessarily lead to a different jet.
The main reason why the jet dynamics is not substantially changed, at least in the early stages of the jet formation and 
propagation, is that the magnetic field structure remains very similar in the innermost disk regions
(see Fig.~\ref{fig::toy_loops_disk}).
This is actually the field structure that is responsible for launching the strongest - and also collimated - jet component.

On the other hand, the dynamo-inefficient zones lead to a different disk mass distribution (see
Fig.~\ref{fig::toy_accr_ej}), which 
naturally affects the evolution of the whole disk-jet system.
In particular, we observe a more turbulent configuration of the poloidal magnetic field (see Fig.\ref{fig::toy_jet_speed}), 
which leads to the ejection of a slower and less massive jet (i.e. with smaller ejection rate, as shown in the bottom panel of Fig. \ref{fig::toy_accr_ej}).
The latter has been proposed already by \citet{2006ApJ...651..272F}.

Another observable is the jet collimation as an imprint of the overall jet dynamics.
There are several options how to best define jet collimation
For example, in \citet{2006ApJ...651..272F,2006MNRAS.365.1131P,2012ApJ...757...65S} the degree of collimation 
has been computed as the ratio of 
the (normalized) mass fluxes in the axial and in the lateral direction, respectively.
Another option is the pure opening angle.
Here we choose a different way to measure the jet collimation quantitatively, taking advantage of the spherical 
coordinates we applied.
More specifically, we compute the opening angle of the jet flow for which the jet has its maximum velocity (or mass flux).
Comparing the angle obtained for different (spherical) radii we obtain a gradual change that in particular demonstrates
the {\em process of collimation}.

What we find from our dynamo simulations is essentially that the jet degree of collimation shows only a weak 
dependence on the strength of the dynamo component $\alpha_\phi$.
This is maybe expected as we know that collimation depends on the profile of the disk magnetic field rather than 
its strength \citep{2006ApJ...651..272F,2006MNRAS.365.1131P}. 
Therefore, no significant differences are found in the jet collimation for a substantially isotropic dynamo, 
while an anisotropic dynamo in general leads to a lower degree of jet collimation 
(see Fig.~\ref{fig::toy_jet_speed}, left).

Another feature that impacts the degree of jet collimation is the presence of magnetic islands, respectively 
magnetized vortices.
This loops severely disturb of the accretion-ejection structure, enhance the turbulence in the outflow flow, 
and also affect the efficiency of mass ejection.

The toroidal magnetic field, which plays a leading role in the jet collimation, is affected by $\alpha_R$ and
$\alpha_\theta$.
In particular, the existence of zones where the mean-field dynamo does not work efficiently, leads to a more 
turbulent configuration 
of both the poloidal and toroidal magnetic field (see Fig.~\ref{fig::toy_jet_speed}, right).
Note, however, that jets also self-generate a substantial toroidal field that usually supports collimation
\citep{1982MNRAS.199..883B}.
Here, the turbulent injection and the turbulent field structure hinder a regular jet toroidal field.
Thus, a weak or non-isotropic dynamo will produce a less collimated jet (see again Fig.~\ref{fig::toy_jet_speed}, right).
To summarize, the dynamo-inefficient zones lead to a more turbulent evolution of both the magnetic field and 
the hydrodynamical quantities, resulting in a more turbulent and less collimated jet structure.

%---------------------------------------------------------------------------------------------------------
\subsection{Early evolution}
\label{sec::toy_early}
Since the target of this toy model is to investigate the effects of the different dynamo components on the 
launching process, we now discuss the impact of the tensor component $\alpha_\theta$ in more detail.
This mostly relates to the very initial evolution of the simulation.

A first result is that for $0<\psi<3$ the evolution of the disk-jet system shows no difference when compared 
with a scalar dynamo.
A likely explanation we find come directly from the choice of the initial configuration of the magnetic field 
in combination with the induction equation.
Since the seed field is purely radial, there is no $B_\theta$-component that can be coupled by a dynamo process.
Therefore, in the initial evolutionary states no contribution can be provided from $\alpha_\theta$.
As the system evolves, the diffusive quenching takes place quite rapidly, leading to a quasi-steady state.
Eventually, the dynamo effects are counterbalanced by magnetic diffusivity and the component $\alpha_\theta$ 
plays a minor role, just because they are weak and had no time to evolve.

However,  when increasing $\alpha_\theta$, as for simulation run {\em th\_A}, its dynamo effect on the temporal 
evolution becomes stronger.
The most important difference to the scalar dynamo simulations is the formation of multiple dynamo-inefficient zones 
within the accretion disk.
As the magnetic field can be amplified only between the dynamo-inefficient zones, this further leads to multiple 
regions in the disk where the magnetic diffusivity does not grow (see Fig.~\ref{fig::toy_dynnum}).

The reason why the early temporal evolution is mostly dominated by the other two dynamo components, essentially 
depends on the initial magnetic field configuration.
On one hand this might look unphysical, as the long-term dynamo amplification of the magnetic field should not 
depend on its initial structure.
On the other hand, a weak field seed must be present in order to initialize a mean-field dynamo effect.

Essentially, a toroidal initial magnetic field leads to the same results (see also \citealt{2014ApJ...796...29S}). 
Similar to the case of a radial initial field, the component $\alpha_\theta$ is not involved in the initial 
temporal evolution
of the $B_{\phi}$, and therefore is able to play a role only when the magnetic field has already saturated.
Thus, the field evolution generated from a purely toroidal initial field leads to results similar to those obtained
from a radial seed field.

This is in contrast to simulations starting from a vertical seed field.
We find a strong impact on the evolution of the system because of the strong shear between the rotating disk and the 
non-rotating (at $t = 0$) corona \citep{2018ApJ...855..130F}.
In addition, this is amplified by the $\alpha_\theta$ dynamo effect of the magnetic field.

This can be nicely seen by our simulation {\em th\_B} applying a vertical seed magnetic field that is derived
from a constant vector
potential $A_\phi = 10^{-5}$ and is applying an anisotropic dynamo with $\psi = 0.1$.
Here, the vertical initial field is able to affect, through the mean-field dynamo, the magnetic field evolution 
and amplification.
A dynamo-inefficient zone is formed around $R\simeq15$.
A collimated outflow is launched, although the overall jet structure shows less collimation 
compared to the simulation with isotropic dynamo (with radial initial field).

%==============================================================================
%==============================================================================
%==============================================================================

\section{Conclusions}
\label{Sec:conclusions}
We have presented MHD dynamo simulations in the context of large-scale jet launching.
Essentially, a magnetic field that is amplified by a mean-field disk dynamo, is able to drive a high speed jet.
All simulations have been performed in axisymmetry, treating all three vector components for the magnetic field 
and velocity.
We have applied the resistive code PLUTO 4.3 \citep{2007ApJS..170..228M}, 
however extended by implementing an additional term in the induction equation
that considers the mean-field dynamo action.

Extending our previous works on mean-field dynamo-driven jets \citep{2014ApJ...796...29S, 2018ApJ...855..130F}, 
here we have essentially investigated the effects of a {\em non-scalar} dynamo tensor.
We have applied

(i) various (ad-hoc) choices for the dynamo tensor components (this paper, paper I), but also 

(ii) an analytical model of turbulent dynamo theory \citep{1995A&A...298..934R} that incorporates both the magnetic 
diffusivity and the turbulent dynamo term, connecting their module and anisotropy by only one parameter, the 
Coriolis number $\Omega^*$ (see paper II, \citealt{Mattia2020}).

In particular we have obtained the following results:

1) We have disentangled different effects of the dynamo tensor components concerning the magnetic field 
   amplification and geometry.
   We find that the strength of the amplification is predominantly related to the dynamo component $\alpha_\phi$.
   The stability of the disk and the launching process can be affected by re-connection events. 
   The field geometry that is favouring re-connection is mainly governed by the dynamo components $\alpha_R$ 
   and $\alpha_\theta$.

2) We find that the component $\alpha_\phi$ is strongly correlated to the amplification of the poloidal magnetic 
field, such that a stronger $\alpha_\phi$ results in a more magnetized disk, 
which then launches a faster, more massive and more collimated jet.
In contrast, the amplification of the poloidal field depends substantially on the existence 
of dynamo-inefficient zones, which, subsequently, affect the overall jet-disk evolution, thus accretion and ejection.

3) We find that not only a stronger dynamo component $\alpha_\theta$ but also a radial component $\alpha_R$ defined
by $\phi < 0.8\lor\phi>1.5$,
respectively, leads to the formation of dynamo-inefficient zones.
The formation of the dynamo-inefficient zones can also be triggered by a vertical component of the 
initial magnetic field, even for a weak dynamo component $\alpha_\theta$.
A strong $\alpha_\theta$ component triggers the formation of the dynamo-inefficient zoned predominantly in the inner 
disk region.
Those loops in general lead to a different evolution of the disk dynamics, since
these zones are dynamo-inefficient and prevent accretion of material from the outer regions of the accretion disk to 
the inner disk that looses mass by accretion and ejection.
      
4) We have investigated how the action of the three different dynamo components affect the jet structure, respectively.
   We find the strength of the magnetic field has a minor influence on the jet speed and mass, however the field 
   geometry, in particular the disk magnetic field profile matters a lot. 
   For lower $\alpha_\phi$ or in presence of dynamo-inefficient zones within the accretion disk, the magnetic 
   field follows a different 
   configuration (with more large-scale magnetic compared a more turbulent structure), which immediately affects
   the jet structure and collimation.
    
5) We have disentangled a clear correlation between the anisotropy of the dynamo tensor and the large-scale
   motion of the jet.
   In particular, dynamos working with a larger $\alpha_\phi$ produce a magnetic field that is able to drive 
   faster jets.
   The reason is that these dynamos lead to a stronger disk magnetization, thus provide more magnetic energy for 
   launching.
   This result nicely couples to correlations between the disk magnetization and various parameters 
   of the jet dynamics as found by \citet{2016ApJ...825...14S}.

6) We have investigated the formation of co-called {\em dynamo-inefficient zones} within the accretion disk and 
   their effect on the disk-jet connection. 
    In particular, such zones are related to a toroidal field reversal with zero derivative, which leads to the 
    formation of multiple loops in the disk. 
    As a consequence, the poloidal magnetic field (in both the disk and the jet) follows a more turbulent 
    evolution, forming e.g. reconnecting magnetic loops, which affects the overall jet launching, 
    the jet mass loading and, subsequently the jet propagation.
    These zones result from certain conditions for the dynamo action, i.e. certain combinations of the dynamo 
    tensor components.

So far we have looked for non-isotropic dynamos applying (ad-hoc) choices for the dynamo tensor components.
In our follow-up paper (paper II), we will apply an analytical model of turbulent dynamo theory \citep{1995A&A...298..934R} 
that incorporates both the magnetic diffusivity and the turbulent dynamo term, connecting their module and anisotropy by 
only one parameter, the Coriolis number $\Omega^*$.

%--------------------------------------------------------------------------------

\acknowledgements
We thank Andrea Mignone and the PLUTO team for the possibility to use their code.
All the simulations were performed on the ISAAC cluster of the Max Planck Institute for Astronomy.
We acknowledge many helpful comments by an unknown referee that lead to a clearer presentation of our results.

%--------------------------------------------------------------------------------
\appendix

\section{Test simulations and comparison to the Literature}
In order to validate our implementation of the mean-field dynamo tensor in the latest version
of PLUTO, we have performed comparison simulations to the reference simulation of \citet{2018ApJ...855..130F},
now restricted to one hemisphere.

Note that while in \citet{2014ApJ...796...29S,2018ApJ...855..130F} the dynamo term was simply coupled with 
the magnetic diffusivity,
here, because of its hyperbolic nature, the $\alpha$-tensor is coupled with the standard hyperbolic MHD 
flux terms, with a correction due to the solenoidal condition of the magnetic field.
Some minor differences in the magnetic field evolution seem to arise from the different implementation 
schemes, 
however, the overall evolution of the system shows very small differences in the strength of 
the physical processes at work.

\begin{figure}
\centering
\includegraphics[width=0.24\textwidth]{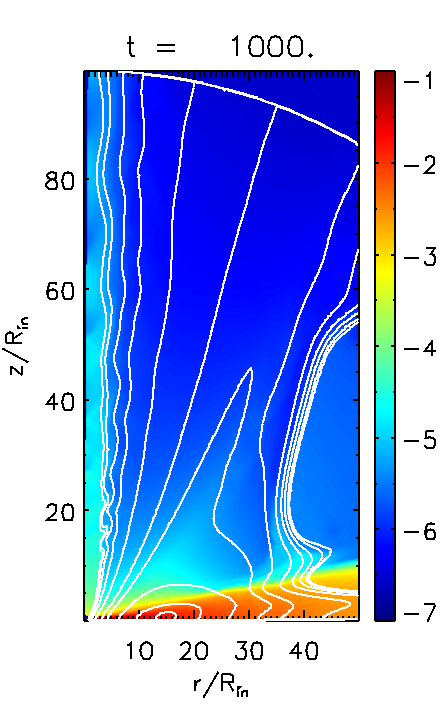}%
\includegraphics[width=0.24\textwidth]{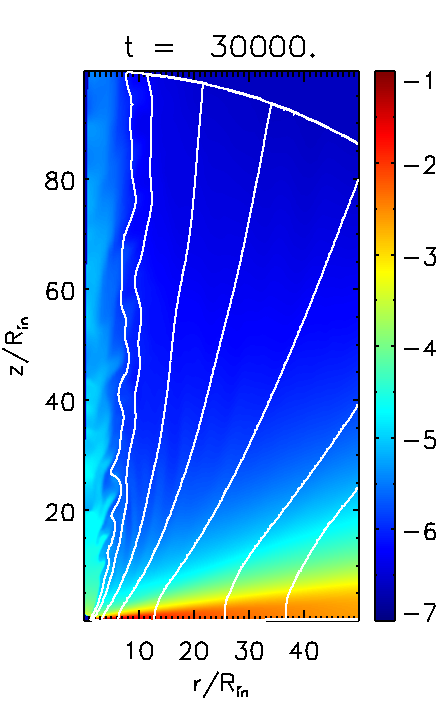}
\caption{Comparison simulation.
Snapshots at $t = [1000,30000]$ for a simulation with the the reference parameters of \citet{2018ApJ...855..130F}, 
now performed with PLUTO 4.3. 
The color map shows the density while the white lines are contours of the vector potential (poloidal magnetic field lines).
}
\label{fig:scalar_0}
\end{figure}

Our simulation runs till $t = 30000$, corresponding to $\simeq 5000$ inner disk rotations.
Figure \ref{fig:scalar_0} shows the evolution of the density and of the magnetic field lines.
We may distinguish three different zones of evolution --  the innermost disk, the outer disk, and  the corona.
The temporal evolution is in very good agreement with \citet{2018ApJ...855..130F}, evolving the same features.

Throughout the inner disk region the magnetic field lines have the typical 
open field lines inclined with respect to the disk surface.
This configuration is particularly favorable for a Blandford-Payne-driven outflow.
The outer disk region is filled with magnetic loops, which are pushed outwards 
by the magnetic pressure gradient and thereby diffusing
through the disk until it is filled with magnetic energy and a {local} 
steady state is reached.

In difference to \citet{2018ApJ...855..130F} we find that the poloidal magnetic energy saturates towards a somewhat level, 
but this is simply because our computational domain is smaller.
Integrated over the whole disk \citet{2018ApJ...855..130F} find a saturation magnitude of $\simeq 2\times10^{-3}$ (in code units), 
while here we reach a saturation value of $\simeq 1.2\times10^{-3}$ (assuming that the lower hemisphere follows
the same evolution as the upper hemisphere).

On the other hand, the accretion and ejection rates saturate at similar magnitude, and also the accretion-ejection ratio agrees
with our previous studies \citep{2018ApJ...855..130F}.
This again strongly supports our conclusion that the different implementation schemes are identical.

%--------------------------------------------------------------------------------------------------------------
\subsection{The dynamo number}
\begin{figure}
\centering
\includegraphics[width=0.24\textwidth]{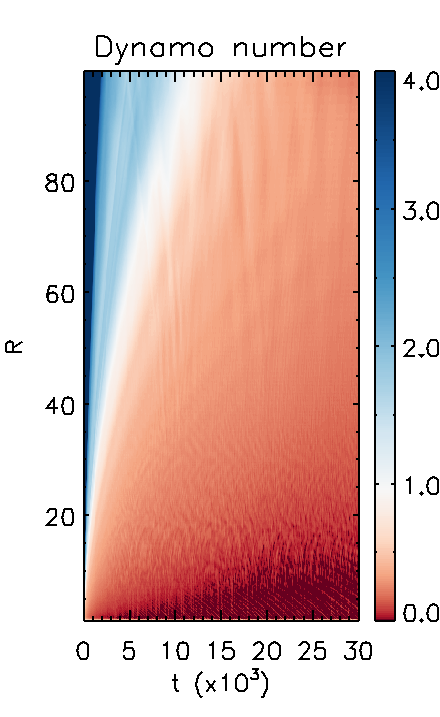}%
\includegraphics[width=0.24\textwidth]{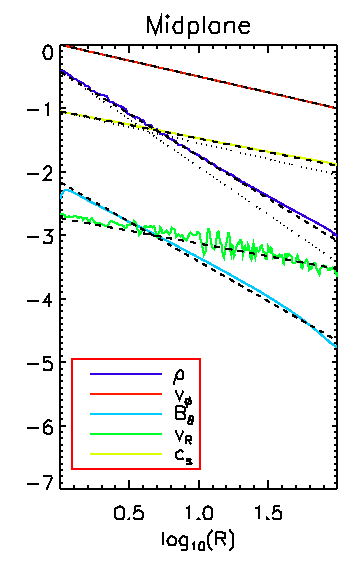}
\caption{Comparison simulation.
Evolution of the dynamo number and MHD variables for a simulation of with the the reference parameters of
\citet{2014ApJ...796...29S}.
Shown is the evolution of the dynamo number as a function of time and radius (left) over time,
and the profile of certain physical quantities along the disk mid-plane (right) at $t = 30000$. 
Colored lines indicate different physical quantities, $\rho$ (blue), $\Cs$ (yellow), $v_\phi$ (red), 
$B_\theta$ (cyan), and $v_R$ (green),
while thin dashed lines show the initial power-law distribution. 
The thick dashed lines show the corresponding fit by a power-law. }
\label{fig::app_rho}
\label{fig:scalar_det}
\end{figure}

One way to examine the evolution of the dynamo action is to look at the time evolution of the dynamo number,
Eq.~\ref{eq::dynnum}.
Dynamo quenching limits the dynamo number to a marginally sub-critical magnitude at which the alpha-dynamo
is balanced by magnetic diffusivity. 
We point out that the critical dynamo number is not known {\it a priori}, but had to be derived from comparison of
parameter studies.
Furthermore, it tells us whether a particular disk region has reached a quasi-steady state.
The evolution of the dynamo number is shown in Figure~\ref{fig::app_rho} as a function of time and radius, respectively.

At $t = 0$ the dynamo number is almost infinite because of the weak magnetization, 
then it decreases starting from the inner radii and then reaching a quasi-steady state also in the outer regions. 
At $t = 5000$ we can distinguish two areas in the profile of the dynamo number. 
For $R < 60$, diffusive quenching has already lead to a field saturation.
For larger radii the diffusivity still decreases as the toroidal magnetic field has not entirely engulfed 
the accretion disk.

At time $t = 15000$ the dynamo-generated magnetic loops are diffused to large radii and the whole system has 
reached a stable configuration.
For all radii the dynamo number is somewhat below 6, which we consider as the critical dynamo number for this 
simulation setup.
We note that this magnitude is similar to  what \citet{2005PhR...417....1B} have suggested, although the critical
dynamo number depends on the geometry and other physical details of the simulation setup.
Going even further in time we see no difference in the temporal evolution nor in the dynamo number.
Thus, all simulations were performed, if not specified otherwise, till $t = 10000$.

%-----------------------------------------------------
\subsection{Mid-plane quantities}
Figure~\ref{fig:scalar_det} shows the distribution of specific physical quantities along the disk mid-plane, measured
at $t = 30000$.
This allows to compare our test simulations to the reference simulation of \citet{2014ApJ...796...29S}.

We have again fitted the simulation data points with a power law in order to extrapolate the power law index $\beta_X$
and compare it with the radial distribution at $t = 0$.
We find that the disk rotation remains Keplerian with $\beta_{v_\phi} = -1/2$.
However, the radial profile of the density distribution changes substantially from $\beta_\rho = -3/2$ to 
$\beta_\rho = -4/3$ 
up to $R\simeq30$, while for larger radii the power index is $\beta_\rho = -5/4$.
As the total mass flux is conserved, the ejection of matter immediately changes the accretion rate over the disk and
is thus related to the changes in the profiles of the mass fluxes.
The radial (accretion) velocity follows a power law index $\beta_{v_R} = -2/5$.
Since we are reaching a longer run time than \citet{2014ApJ...796...29S}, 
we are now able to get rid of the oscillations and also the reversal found by \citet{2014ApJ...796...29S} in the outer 
disk regions (as their magnetic field was not yet diffused across the whole accretion disk).

The power-law coefficient of the sound speed changes during $t = 0$ and $t = 10000$ from $\beta_{\rm{cs}} = -1/2$ 
to $\beta_{\rm{cs}} = -3/7$, 
which tells us that the mean-field dynamo only slightly changes its strength as due to the disk sound speed 
through the disk-jet evolution
(see Eq.~\ref{eq::dynamo}).
This change does not lead to any strong net effect on the temporal evolution of the disk-jet system, 
therefore we again find difference to our previous results \citep{2014ApJ...796...29S, 2018ApJ...855..130F}.

Also the angular magnetic field component $B_\theta$ follows the same power law, namely $\beta_{B_\theta} = -5/4$.
Note, however, that we do not find the decrease in the outer disk regions ($R\geq40$) as found
in \citet{2014ApJ...796...29S},
simply because of our longer simulation time.

Overall, by quantifying essential dynamical properties of our simulation results, we find perfect agreement with the 
previous results that are based on a numerically different implementation of magnetic diffusivity and mean-field dynamo. 

%---------------------------------------------------------------------
\section{Control volumes and fluxes}
\label{sec::integrals}
Here we define how we integrate global quantities that are used throughout the paper.
The accretion rate is calculated by integrating the net radial mass flux through the disk, defined 
by an opening angle $\theta_S \equiv \arctan(2H/r)$,
\begin{equation}
    \dot{M}_{\rm{acc}}(R) = 2\pi R \int_{\pi/2}^{\pi/2 - \theta_S}  \rho v_R R d\theta,
\end{equation}
while the ejection rate is calculated integrating the outflow in vertical direction (through the disk surface),
\begin{equation}
  \dot{M}_{\rm{eje}}(R; \theta_S) = \int_{R_{\rm{in}}}^{R} \rho v_{\theta}(\tilde{R}) 2\pi \tilde{R} d\tilde{R},
\end{equation}
respectively.
The magnetic disk energy (poloidal or toroidal) is integrated from a radius of choice $R$ to the outer radius
$R_{\rm{out}}$, and from the disk midplane to the disk surface, defined by $\theta_S$. 
We thus consider the disk magnetic energy outside $R$ for our considerations,
\begin{equation}
    E_{\rm{mag}} = \int_{R}^{R_{\rm{out}}} \int_{\pi/2-\theta_S}^{\pi/2}  \frac{1}{2} B^2 \sin(|\theta|) 2\pi R^2d\theta dR. 
\end{equation}
The so-called {\em disk magnetic field} (and also the {\em disk magnetization}) is simply calculated as the average 
value of the magnetic field, at each radius, within the initial disk defined by $\theta_i \equiv \arctan(H/r)$,
\begin{equation}
    B_{\rm{disk}}(R) = \DS\frac{1}{\theta_i}  \int_{\pi/2-\theta_i}^{\pi/2} B(R, \theta) d\theta \\
\end{equation}
while the so-called {\em disk diffusivity} is the average value of the diffusivity at a certain radius within 
the initial accretion disk,
\begin{equation}
    \eta_{\rm{disk}}(R,t) = \frac{1}{\theta_i} \int_{\pi/2-\theta_i}^{\pi/2} \eta(R, \theta) d\theta
\end{equation}

%==============================================================================
%
%
%
%==============================================================================
\vspace{2mm}
% THE REFERENCES ARE HERE
\bibliographystyle{apj}
% \bibliography{dg_Literatur}
 %\bibliography{main}

%---------------------------------------------------------------
\end{document}